\documentclass[twocolumn,amsmath,amssymb,pra]{revtex4-2}
\usepackage{txfonts}
\usepackage{dcolumn}% Align table columns on decimal point
\usepackage{color}
\usepackage{graphicx}
\usepackage{adjustbox}
\usepackage{rotating}
\usepackage{array}
\usepackage{caption}

\usepackage{xr}
\usepackage{nicematrix} % for boxed blocks
\usepackage{tikz}       % needed for drawing

\usepackage{amssymb}
\usepackage{amsmath}
\usepackage{mathrsfs}
\usepackage{braket}
\usepackage{bm}
\usepackage{bbm}
\usepackage{mathtools}

\usepackage[
  pdfauthor={Chun Hei LEUNG},
  pdftitle={Undamped Modes in an N-Qubit Heisenberg Chain with Collective Dissipation},
  pdfkeywords={arxiv v1},
  colorlinks=true, linkcolor=blue, urlcolor=blue, anchorcolor=blue, citecolor=red, bookmarksnumbered
]{hyperref}

\usepackage{lipsum}

\begin{document}
\title{\textbf{Undamped Modes in an N-Qubit Heisenberg Chain with Collective Dissipation} 
}% 

\author{Chun Hei Leung$^1$}

\affiliation{
$^1$School of Physics and Astronomy and Centre for the Mathematics and Theoretical Physics of Quantum Non-equilibrium Systems, University of Nottingham, Nottingham NG7 2RD, United Kingdom
}

\date{\today}% It is always \today, today,

\begin{abstract}

   We investigate the undamped behaviors in a spin-1/2 Heisenberg chain coupled with an environment via collective spin jump operators. Using the Bethe ansatz basis, we show that undamped modes exist for any chain length $N \geq 3$. These modes remain robust against variations in the system parameters, including the specific form of the collective dissipation, and the external field. Exploiting the Bethe ansatz solution, we further characterize the number of undamped modes and their oscillation frequencies, uncovering long-lived coherent dynamics in open integrable quantum systems.
  
% \begin{description}
% Secondary publications and information retrieval purposes.
% You may use the \texttt{description} environment to structure your abstract;
% use the optional argument of the \verb+\item+ command to give the category of each item. 
% \end{description}
\end{abstract}

\maketitle

%\tableofcontents

%=============================================================================

\section{Introduction}

    Open quantum systems are unavoidably coupled to their environments, and this interaction typically leads to dissipation and decoherence. Consequently, quantum coherence is gradually degraded and information encoded in the system leaks into uncontrolled environmental degrees of freedom. In most cases, the long-time dynamics relax toward a steady state~\cite{10.1093/acprof:oso/9780199213900.001.0001, Spohn1977}, with coherent behavior surviving only transiently. Nevertheless, certain systems~\cite{PhysRevLett.122.015701, PRXQuantum.5.030325, bucaNonstationaryCoherentQuantum2019} support undamped modes: dynamical components that remain coherent indefinitely despite ongoing coupling to the environment. These modes identify exceptional sectors of open-system dynamics in which dissipation is ineffective, providing a route to robust quantum control, and long-lived quantum information storage~\cite{Koch_2016, esencan2026timecrystalspassivelyprotected, PhysRevA.57.3276}.

    Standard approaches to preserve quantum information rely on active quantum error correction (QEC), which combines repeated measurements, syndrome detection, and feedback to correct errors caused by noise. Although powerful in principle, active quantum error correction introduces significant practical challenges. It requires continuous external intervention, fast and high-fidelity measurements, additional qubits for redundancy, and real-time control loops. These demands lead to substantial operational overhead, measurement back-action, and scalability issues, particularly in many-body or solid-state platforms. 
    By contrast, passive quantum error correction~\cite{PhysRevLett.79.3306, liu2023dissipativephasetransitionspassive, doi:10.1126/science.1064460} avoids active intervention by encoding information in intrinsically protected structures such as decoherence-free subspaces and noiseless subsystems~\cite{PhysRevLett.81.2594, PhysRevA.63.012301, PhysRevLett.100.030501}. In this setting, robustness emerges from the internal structure of the system and its coupling to the environment. Undamped modes can be understood as a dynamical realization of this idea: they correspond to protected sectors of the Liouvillian spectrum that evade decay and support persistent coherent evolution.

    In this work, we investigate an open quantum system based on the Heisenberg model, a cornerstone of quantum information theory, including applications in quantum state transfer~\cite{PhysRevLett.91.207901, PhysRevLett.92.187902} and the implementation of quantum gates~\cite{2230956.2230965, Nguyen2024Programmable}. It is also central to quantum many-body physics, where it provides insight into collective behaviors in interacting spin systems, such as magnetism~\cite{PhysRev.133.A768} and quantum phase transitions~\cite{PhysRevLett.37.120}. 
    Notably, the one-dimensional Heisenberg model is exactly solvable via the Bethe ansatz~\cite{PhysRev.150.321}, making it a valuable benchmark for analytical and numerical studies. Leveraging this approach, we show that undamped modes can arise in a Heisenberg chain of arbitrary size $N \geq 3$ when coupled to an environment through collective spin jump operators. The Bethe solution further allows us to determine both the number of undamped modes and their oscillation frequencies.
    
\subsection{Conditions for the Existence of Undamped Modes in Open Quantum System}

    For an open quantum system interacting with a Markovian environment, the time evolution of the system's density matrix $\rho$ is governed by the Lindblad master equation~\cite{Lindblad1976OnTG, 10.1063/1.522979, PhysRevB.102.115109},
    \begin{eqnarray}
        \frac{d}{dt} \rho = \mathscr{L} [\rho] 
            &\equiv& -\mathbbm{i} [ H , \rho ] + \sum_{i} \gamma_{i} \, \mathcal{D}_{i} [\rho] ,
    \label{eq: master equation}
    \end{eqnarray}
    where $\mathscr{L}$ denotes the Liouvillian superoperator and $H$ is the system Hamiltonian. The dissipators
    \begin{eqnarray}
        \mathcal{D}_i [\rho] 
            &\equiv& L_{i}^{\phantom{\dag}} \rho L_{i}^{\dagger} - \frac{1}{2} L_{i}^{\dagger} L_{i}^{\phantom{\dag}} \rho - \frac{1}{2} \rho L_{i}^{\dagger} L_{i}^{\phantom{\dag}} .
        % \mathcal{D}_i [\rho] 
        %     &\equiv& L_{i} \rho L_{i}^{\dagger} - \frac{1}{2} L_{i}^{\dagger} L_{i} \rho - \frac{1}{2} \rho L_{i}^{\dagger} L_{i} .
    \end{eqnarray}
    capture the non-unitary effects due to the environment and are defined in terms of the jump operators $L_{i}$ with corresponding non-negative damping rates $\gamma_{i} \geq 0$.

    The solution of Eq.~\eqref{eq: master equation} can be expressed~\cite{PhysRevA.98.042118, longhiQuantumMpembaEffect2025, PhysRevLett.125.060601} as
    \begin{eqnarray}
        \rho (t) 
            &=& \sum_{j} c_{j} \, e^{\Lambda_{j} \, t} r_{j} ,
    \label{eq: eigenmodes expansion}
    \end{eqnarray}
    where $c_j \equiv \operatorname{Tr} [l^{\dagger}_{j} \, \rho (0)] $ is the initial projection amplitude onto the $j$-th (right) eigenmode. $\Lambda_j$, $r_j$, and $l_j$ are the eigenvalues, right eigenmodes, and left eigenmodes of the Liouvillian superoperator, respectively. 
    The Liouvillian eigenvalues $\Lambda_j$ determine the capture the dynamical characteristics of the system, with $\Gamma_{j} \equiv -\operatorname{Re}[\Lambda_{j}]$ specifying the decay rate and $\omega_{j} \equiv \operatorname{Im}[\Lambda_{j}]$ the (angular) oscillation frequency of the $k$-th eigenmode, whose classification~\cite{PRXQuantum.5.030325, Baumgartner_2008} is summarized in Table~\ref{table: eigenmodes_classification}.
    \begin{table}[hbtp]
        \caption{The classification of Liouvillian eigenmodes.}
        \label{table: eigenmodes_classification}
        \begin{ruledtabular}
            \begin{tabular}{c|cc}
                & 
                $\Gamma_{j} \equiv -\operatorname{Re} [\Lambda_{j}] = 0$ \footnotemark[1] &
                $\Gamma_{j} \equiv - \operatorname{Re} [\Lambda_{j}] > 0$ \footnotemark[2] \\
                \colrule
                $\omega_{j} \equiv \operatorname{Im} [\Lambda_{j}] \neq 0$ & Persistent oscillatory & Underdamped \\
                $\omega_{j} \equiv \operatorname{Im} [\Lambda_{j}] = 0$ & Steady & Overdamped \\
            \end{tabular}
            \begin{minipage}[t]{0.45\linewidth}
                \footnotetext[1]{Undamped modes}
            \end{minipage}
            \hfill
            \begin{minipage}[t]{0.45\linewidth}
                \footnotetext[2]{Damped modes}
            \end{minipage}
        \end{ruledtabular}
    \end{table}
    
    Interactions with the environment typically lead to dissipation in a quantum system, causing it to relax toward a steady state and progressively damping any coherent oscillatory behavior over time. Remarkably, however, under certain conditions~\cite{Buca_2012, PhysRevA.77.052301, Zhang_2020, Booker_2020, albertSymmetriesConservedQuantities2014, PhysRevLett.125.240405}, the system can sustain persistent coherent dynamics that survive indefinitely, despite the presence of environmental interaction.

    In particular~\cite{engineercoherentoscillatorymodes}, it has been shown that undamped eigenmodes of Liouvillian exist when the system Hamiltonian $H$ and the jump operators $L_{i}$ can be expressed in the following block-diagonal form, 
    \begin{eqnarray}
        H = 
            \begin{pmatrix}
                H^{a} & 0 & 0 \\
                0 & H^{b} & 0 \\
                0 & 0 & H^{\mathrm{res}}
            \end{pmatrix}
            _\mathcal{A} 
        ,
        L_{i} = 
            \begin{pmatrix}
                \Xi_{i}^{a} & 0 & 0 \\
                0 & \Xi_{i}^{b} & 0 \\
                0 & 0 & \Xi^{\mathrm{res}}_{i}
            \end{pmatrix}
            _\mathcal{A} ,
    \label{eq: block-diagonal structure}
    \end{eqnarray}
    in an appropriate orthonormal basis $\mathcal{A}$.
    Here $H^{a}$ and $\Xi_{i}^{a}$, $H^{b}$ and $\Xi_{i}^{b}$, as well as $H^{\mathrm{res}}$ and $\Xi^{\mathrm{res}}_{i}$ are all square matrices, with each pair having the same dimension.
    Under these conditions, (trivial) steady eigenmodes arise naturally within each invariant subspace (each block), taking the form of
    \begin{eqnarray}
            \begin{pmatrix}
                R^{a} & 0 & 0 \\
                0 & 0 & 0 \\
                0 & 0 & 0
            \end{pmatrix}
            _\mathcal{A}
            ,
            \begin{pmatrix}
                0 & 0 & 0 \\
                0 & R^{b} & 0 \\
                0 & 0 & 0
            \end{pmatrix}
            _\mathcal{A} 
        \text{, and }
            \begin{pmatrix}
                0 & 0 & 0 \\
                0 & 0 & 0 \\
                0 & 0 & R^{\mathrm{res}}
            \end{pmatrix}
            _\mathcal{A} ,
    \end{eqnarray}
    where $R^{a}$, $R^{b}$ and $R^{\mathrm{res}}$ are all square matrices.
    
    Furthermore, if $\Xi_{i}^{a} = \Xi_{i}^{b}$ for all $i$ and the system Hamiltonian satisfies the ``strong $\Delta_H$ condition'' 
    \begin{eqnarray}
        \Delta_H \equiv H^{a} - H^{b} &=& \omega \, \mathbb{I} , 
    \end{eqnarray}
    then a pair of undamped eigenmodes also exists, taking the form
    \begin{eqnarray}
            \begin{pmatrix}
                0 & 0 & 0 \\
                R & 0 & 0 \\
                0 & 0 & 0
            \end{pmatrix}
            _\mathcal{A}
        \text{ and }
            \begin{pmatrix}
                0 & R & 0 \\
                0 & 0 & 0 \\
                0 & 0 & 0
            \end{pmatrix}
            _\mathcal{A} ,
    \label{eq: oscillatory modes}
    \end{eqnarray}
    where $R$ is a square matrix, 
    with Liouvillian eigenvalues $+ \mathbbm{i} \omega$ and $- \mathbbm{i} \omega$. 
    These correspond to a pair of persistent oscillatory modes when $\omega \neq 0$ with (angular) oscillation frequency $\omega$; when $\omega = 0$, they regress as steady modes.

\subsection{The model}

    The open quantum system under consideration is a spin-1/2 isotropic (XXX) Heisenberg chain~\cite{Heisenberg1928} of $N$ sites with periodic boundary conditions, subjected to a uniform external field. The system Hamiltonian reads 
    \begin{eqnarray}
        H
            &\equiv& H_{J} + H_{h} .
    \end{eqnarray}
    Here, the Heisenberg spin-spin interactions with coupling constant $J$ is given by
    \begin{eqnarray}
        H_{J}
            &\equiv& \sum_{n=1}^{N} J \, \vec{S}_{n} \cdot \vec{S}_{n+1} 
                \!\!\qquad\qquad , \vec{S}_{n+N} \equiv \vec{S}_{n} \\
            &=& \sum_{n=1}^{N} J \, \left( S^{x}_{n} \, S^{x}_{n+1} + S^{y}_{n} \, S^{y}_{n+1} + S^{z}_{n} \, S^{z}_{n+1} \right) ,
            % &=& - \sum_{n=1}^{N} J \left( \frac{1}{2} S^{+}_{n} S^{-}_{n+1} + \frac{1}{2} S^{+}_{n} S^{-}_{n+1} + S^{z}_{n} S^{z}_{n+1} \right) ,\\
    \end{eqnarray}
    and the interaction with the external field $\vec{h} = (h^{x}, h^{y}, h^{z})$ is given by
    \begin{eqnarray}
        H_{h}
            &\equiv& \vec{h} \cdot \vec{S} \\
            % &=& \sum_{n=1}^{N} h^{x} S^{x}_{n} + h^{y} S^{y}_{n} + h^{z} S^{z}_{n} \\
            &=& h^{-} \, S^{+} + h^{+} \, S^{-} + h^{z} \, S^{z} ,
    \end{eqnarray}
    where $h^{\pm} = \frac{1}{2} (h^{x} \pm \mathbbm{i} h^{y})$, $\vec{S}_{n} = (S^{x}_{n}, S^{y}_{n}, S^{z}_{n})$ is the local spin operator at site $n$, and $\vec{S} = \sum_{n=1}^{N} \vec{S}_{n}$ is the collective spin operator. The local spin raising and lowering operators are defined as $S^{\pm}_n = S^{x}_{n} \pm \mathbbm{i} S^{y}_{n}$, and the collective spin raising and lowering operator are defined as $S^{\pm} = \sum_{n=1}^{N} S^{\pm}_{n}$.    
    Moreover, the system under consideration is coupled to an environment via a set of jump operators $ L_{i}$, which can be expressed as functions of the collective spin operators,
    \begin{eqnarray}
        L_{i}
            &\equiv& f_{i} \, ( S^{x} , S^{y} , S^{z} ) ,
    \label{eq: nonlinear jump operator}
    \end{eqnarray}
    where $f_{i}$ are (non-commutative) polynomials in the collective spin operators.    
    To keep the discussion tractable, we focus initially on linear collective jump operators of the form
    \begin{eqnarray}
        L_{i}
            &\equiv& \ell^{0}_{i} \, \mathbb{I} + \ell^{x}_{i} \, S^{x} \:\! + \ell^{y}_{i} \, S^{y} \:\! + \ell^{z}_{i} \, S^{z} \\
            &=& \ell^{0}_{i} \, \mathbb{I} + \ell^{-}_{i} S^{+} + \ell^{+}_{i} S^{-} + \ell^{z}_{i} \, S^{z} ,
    \end{eqnarray}
    where $\ell^{\pm} = \frac{1}{2} (\ell^{x} \pm \mathbbm{i} \ell^{y})$. The results can subsequently be extended to the non-linear case.

    The backbone of this open quantum system is the one-dimensional Heisenberg model, a cornerstone of statistical physics renowned for its integrability. Remarkably, all eigenvalues and eigenstates of its Hamiltonian are exactly solvable via the Bethe ansatz, analytically.
    To demonstrate the block-diagonal structure of the system and the presence of undamped or oscillatory modes, we adopt the Bethe (ansatz) basis. While the Clebsch–Gordan (CG) basis could also serve this purpose, it is less convenient for a generic $N$-qubit system.
  
\section{Insight from SU(2) Decomposition of Coupled Spin-1/2 Systems}

    The Heisenberg model possesses a global SU(2) spin-rotation symmetry, reflecting the invariance of the system under simultaneous rotations of all spins. Consequently, in an SU(2) decomposition, the representation splits into distinct spin sectors (multiplets), each corresponding to a specific total spin. These multiplets form invariant subspaces, which appear as blocks along the diagonal in the matrix representation, indicating that states in different sectors do not mix. This block-diagonal structure not only simplifies calculations but also underlies the emergence of degenerate or collective modes within the system.
 
    In group theory, when $N$ spin-$1/2$ particles are coupled together, the irreducible representation of the SU(2) decomposition is given by
    \begin{eqnarray}
        \left( \mathbf{\frac{1}{2}} \right)^{\otimes N} 
            &=& \bigoplus_{\mathbb{S} = \mathbb{S}_{\mathrm{min}}}^{\mathbb{S}_{\mathrm{max}}} (\mathbb{S})^{\oplus \, d_{\mathbb{S}}} ,
    \end{eqnarray}
    where $\mathbb{S}$ represents the total spin quantum number, ranging from $\mathbb{S}_{\mathrm{min}}$ to $\mathbb{S}_{\mathrm{max}}$, with
    \begin{eqnarray}
        \mathbb{S}_{\mathrm{max}} = \frac{N}{2}, 
        \quad
        \mathbb{S}_{\mathrm{min}} = 
        \begin{cases} 
            0 & \text{for even } N \\
            \frac{1}{2} & \text{for odd \ } N
        \end{cases} ,
    \end{eqnarray}
    and the multiplicity of spin $\mathbb{S}$ is
    \begin{eqnarray}
        d_{\mathbb{S}} 
        &=& \binom{N}{\frac{N}{2} - \mathbb{S}} - \binom{N}{\frac{N}{2} - \mathbb{S} - 1}.
    \end{eqnarray}
    Table~\ref{table: SU2 decomposition} lists the resulting representations for small $N$.
    \begin{table}[!h]
    \caption{SU(2) Irreducible Decomposition for $N$ qubits.}
    \label{table: SU2 decomposition}
    \begin{ruledtabular}
    \renewcommand{\arraystretch}{1.8} % Increase row height
    \setlength{\tabcolsep}{12pt}      % Increase horizontal padding
    \begin{tabular}{@{\hskip 64pt} r|l @{\hskip 64pt}}
        $N$ & SU(2) Decomposition \\
        \hline
        2 & $\left(\mathbf{0}\right)^{\oplus 1} \;\oplus\; \left(\mathbf{1}\right)^{\oplus 1}$ \\
        3 & $\left(\mathbf{\frac{1}{2}}\right)^{\oplus 2} \;\oplus\; \left(\mathbf{\frac{3}{2}}\right)^{\oplus 1}$ \\
        4 & $\left(\mathbf{0}\right)^{\oplus 2} \;\oplus\; \left(\mathbf{1}\right)^{\oplus 3} \;\oplus\; \left(\mathbf{2}\right)^{\oplus 1}$ \\
        5 & $\left(\mathbf{\frac{1}{2}}\right)^{\oplus 5} \;\oplus\; \left(\mathbf{\frac{3}{2}}\right)^{\oplus 4} \;\oplus\; \left(\mathbf{\frac{5}{2}}\right)^{\oplus 1}$ \\
    \end{tabular}
    \end{ruledtabular}
    \end{table}
    \\ \noindent
    For $N \geq 3$, the decompositions not only split into different total spins but also shows multiple copies of the same spin value. These repeated multiplets form subspaces (blocks) with the same dimension, and provide the necessary structure for constructing the undamped or oscillatory eigenmodes discussed in Eq.~\eqref{eq: oscillatory modes}.

\section{Brief Review of the (Coordinate) Bethe Ansatz}

    The Bethe ansatz, introduced by Hans Bethe in 1931~\cite{Bethe1931Zur}, yields exact eigenvalues and eigenstates for the one-dimensional isotropic Heisenberg model. Since then, numerous related techniques have been developed for exactly solvable quantum many-body problems, all falling under the general umbrella of the Bethe ansatz, including the algebraic Bethe ansatz and analytic Bethe ansatz. The original version introduced by Bethe is now referred to as the coordinate Bethe ansatz, which is the approach we will employ in the following.

    Bethe’s method begins with a reference state (also called the vacuum state), in which all spins are aligned up, 
    \begin{eqnarray}
        \ket{\Omega} &\equiv& \bigotimes_{n=1}^{N} \ket{\uparrow}_{n} .
    \end{eqnarray}
    From this reference state, one can construct states with $M$ excitations by applying local spin-lowering operators $S^{-}_{n}$ at $M$ distinct sites,
    \begin{eqnarray}
        \ket{n_{1}, \cdots, n_{M}} &\equiv& S^{-}_{n_{1}} \cdots S^{-}_{n_{M}} \ket{\Omega} .
    \end{eqnarray}
    The Bethe’s preliminary ansatz expresses the eigenstates of the one-dimensional isotropic Heisenberg Hamiltonian $H_{J}$, known as Bethe states, as
    \begin{eqnarray}
        \ket{\psi} 
            &\equiv& \sum_{\{ n \}} \psi(n_{1}, \cdots, n_{M}) \ket{n_{1}, \cdots, n_{M}} ,
    \end{eqnarray}
    with the Bethe wavefunction defined by
    \begin{eqnarray}
        \psi(n_{1}, \cdots, n_{M})
            \equiv \sum_{\mathcal{P}} \exp \left[ \mathbbm{i} \left( \sum_{j=1}^{M} k_{\mathcal{P}_{j}} n_{j} + \frac{1}{2} \sum_{i<j} \theta_{\, \mathcal{P}_{i} \mathcal{P}_{j}} \right) \right] ,
    \label{eq: bethe wavefunction}
    \end{eqnarray}
    where the sum of $\{ n \}$ runs over the $\binom{N}{M}$ sets of $M$ increasing integers varying from $1$ to $N$, i.e. $1 \leq n_{1} < \cdots < n_{M} \leq N$, and the sum of $\mathcal{P}$ runs over all $M!$ permutations of the labels $\{ 1, \cdots, M \}$.
    The quantities $k_{\mu}$ are referred to as the quasi-momenta, and $\theta_{\mu \nu}$ are the scattering phases, defined by
    \begin{eqnarray}
        e^{\mathbbm{i} \theta_{\mu \nu}} 
            &\equiv& - \frac{e^{\mathbbm{i} (k_{\mu} + k_{\nu})} + 1 - 2 e^{\mathbbm{i} k_{\mu}}}{e^{\mathbbm{i} (k_{\mu} + k_{\nu})} + 1 - 2 e^{\mathbbm{i} k_{\nu}}} . 
    \label{eq: scattering phase} 
        %% \\
        %% \Leftrightarrow 2\cot[\frac{\theta_{\mu \nu}}{2}] 
        %%     &\equiv& \cot[\frac{k_{\mu}}{2}] - \cot [\frac{k_{\nu}}{2}] .
    \end{eqnarray}
    Special care must be taken when $k_{\mu} = k_{\nu}$, where a singularity arises; by convention, $\theta_{\mu \nu} = 0 $ in this case. 
    For illustration, the 1-magnon states ($M=1$) can be given explicitly as 
    \begin{eqnarray}
    \ket{\psi} 
        = \sum_{n_{1}=1}^{N} e^{\mathbbm{i} k_{1} n_{1}} \ket{n_{1}} ,
    \end{eqnarray}
    and the 2-magnon states ($M=2$) take the form 
    \begin{eqnarray}
    \ket{\psi} 
        = \sum_{\substack{n_1, n_2 = 1 \\ \{ n_1 < n_2 \} }}^{N} \left( e^{\mathbbm{i} (k_{1} n_{1} + k_{2} n_{2} + \frac{1}{2} \theta_{12})} + e^{\mathbbm{i} (k_{1} n_{2} + k_{2} n_{1} + \frac{1}{2} \theta_{21})} \right) \ket{n_{1}, n_{2}} .
    \end{eqnarray}

    A Bethe state with an arbitrary set of quasi-momenta $\{ k_{\mu} \}$ is not necessarily an eigenstate of the isotropic Heisenberg Hamiltonian $H_{J}$. Only those states for which the set of quasi-momenta satisfies the Bethe equation (which arises from imposing periodic boundary condition), 
    \begin{eqnarray}
        N k_{\mu}
            &=& 2 \pi I_{\mu} + \sum_{\nu \neq \mu} \theta_{\mu \nu} , \quad \forall \mu,
    \end{eqnarray}
    where the quantities $I_{\mu}$ (taking values as integers or half-integers) are called the Bethe quantum numbers, correspond to true eigenstates of $H_{J}$. Such states are referred to as on-shell Bethe states, and their corresponding eigenvalues (eigenenergies) are  
    \begin{eqnarray}
        E_{J} 
            &=& J \, \frac{N}{4} - J \sum_{\nu = 1}^{M} (1 - \cos [k_{\nu}]) .
        \label{eq: E_J}
    \end{eqnarray}
    In addition, on-shell Bethe states are simultaneous eigenstates of $S^{z}$ and $S^{2} \equiv (S^{x})^2 + (S^{y})^2 + (S^{z})^2$, since $[ H_{J} , S^{2} ] = [ H_{J} , S^{z} ] = [ S^{2} , S^{z} ] = 0 $.

\subsection{Solutions to the Bethe Equation}

    The Bethe equation simplifies an exponentially large problem into algebraic equations that scale linearly with magnons, making it possible to obtain exact results and probe the physics of large systems.

    Solving the Bethe equations would yield the so-called highest-weight states~\cite{PhysRevE.88.052113}, whose set of quasi-momenta $\{ k_{\mu} \}$ does not include zero quasi-momentum, and which are annihilated by the collective spin raising operator $S^{+}$, i.e. $S^{+} \ket{\psi} = 0$, thereby possess the maximal $S^{z}$ within their multiplets, as the name suggests. 
    However, the highest-weight states alone do not form a complete solution set. The remaining solutions can be generated by repeatedly applying the collective spin lowering operator $S^{-}$ to the highest-weight states~\cite{PhysRevE.88.052113}, these are referred to as descendant states of the associated highest-weight state. This process terminates upon reaching the lowest-weight states, which are annihilated by $S^{-}$.     
    If a descendant state is obtained by applying $S^{-}$, $\zeta$ times, to a highest-weight state with quasi-momenta $\{ k_{1}, \cdots, k_{M} \}$, then the quasi-momenta of the descendant state are obtained by appending zero quasi-momentum entries $\zeta$ times, i.e. $\{ k_{1}, \cdots, k_{M} , k_{M + 1} = 0, \cdots, k_{M + \zeta} = 0 \}$. 
    Thus, all descendant states share the same $E_{J}$ as their associated highest-weight state, in accordance with Eq.~\eqref{eq: E_J}.
    
    For two on-shell Bethe states, $\ket{D}$ and $\ket{A}$, if $S^{-} \ket{A} \sim \ket{D}$ (or equivalently $S^{+} \ket{D} \sim \ket{A}$), then $\ket{D}$ is called the immediate descendant of $\ket{A}$, and $\ket{A}$ is called the immediate ancestor of $\ket{D}$.
    The collective spin quantum number of a highest-weight state is given by $s = \frac{N}{2} - M$ and all its associated descendants share the same value of $s$. The collective spin projection quantum number $m$ of a highest-weight state is equal to its collective spin quantum number, $m = s$, as it has the maximal $S^{z}$; while it decreases as $m = s - \zeta$, with $\zeta \in \{ 0, \cdots, 2s \} $ for its descendant which obtained after $\zeta$ applications of $S^{-}$.
    
\subsection{The Bethe Basis}

    The on-shell Bethe states, including both the highest-weight states and their descendant states, form a complete basis and are mutually orthogonal~\cite{1989.Slavnov.TMP.79}. However, they are not automatically normalized, as defined in Eq.~\eqref{eq: bethe wavefunction}.
    
    The norm of the highest-weight states can be calculated using the determinant of the Gaudin matrix $G$~\cite{Gaudin_2014, Korepin1982Calculation}, which allows for proper normalization.
    \begin{eqnarray}
        \sum_{\{ n \}} \| \psi(n_{1}, \cdots, n_{M}) \|^2          
            &=& \det[G] ,
    \end{eqnarray}
    where
    \begin{eqnarray}
        G 
            &\equiv& 2 \pi \frac{\partial (I_{1} \cdots I_{M})}{\partial (k_{1} \cdots k_{M})} , 
        \\
        2 \pi \frac{\partial I_{\mu}}{\partial k_{\nu}} 
            &=& \begin{cases}
                N - \substack{ \sum_{\xi = 1}^{M} \\ \{ \xi \neq \mu \} } \frac{4 (1 - \cos[k_\xi])}{(2 - e^{-i k_\mu} - e^{+i k_\xi}) (2 - e^{-i k_\xi} - e^{+i k_\mu})}
                \ \text{ for } \nu = \mu
                \\
                \phantom{N - \substack{ \sum_{\xi=1}^{M} \\ \{ \xi \neq \mu \} }}
                \frac{4 (1 - \cos[k_\mu])}{(2 - e^{-i k_\mu} - e^{+i k_\nu}) (2 - e^{-i k_\nu} - e^{+i k_\mu})}
                \ \text{ for } \nu \neq \mu
            \end{cases} 
    \end{eqnarray}
    The norm of descendant states generally cannot be computed directly using the Gaudin determinant, due to singularities that arise in the scattering phase, Eq.~\eqref{eq: scattering phase}, when multiple zero quasi-momenta are present. However, their norms can be deduced using the formula:
    \begin{eqnarray}
        (S^{-})^{\zeta} \ket{s, m} 
            = \sqrt{\frac{(s+m)! (s-m+\zeta)!}{(s-m)!(s+m - \zeta)!}} \ket{s, m - \zeta}.
    \end{eqnarray}

    At this stage, once the Bethe ansatz solutions are obtained, the Bethe basis $\mathcal{B}$ can be constructed accordingly. It provides the complete orthonormal basis needed to reveal the block-diagonal structure in Eq.~\eqref{eq: block-diagonal structure}. Although its explicit expression is not necessary for the subsequent derivation, it is presented in the Appendix~\ref{bethe basis tables} for $N=3,4,5$, to provide a complete and transparent reference.
    
\section{Existence of Undamped Modes}
    
    Using the properties of the Bethe basis states, it is easy to see that both the system Hamiltonian $H$ and the jump operators $L_{i}$ exhibit a block-diagonal structure, with each block corresponding to a multiplet. This arises because they are composed entirely of $\mathbb{I}, H_{J}, S^{z}, S^{+}, S^{-}$, which only connect Bethe states within the same multiplet. Their relevant non-zero matrix elements are as follows.
    % \\ \noindent
    \linebreak
    Since the Bethe basis states are mutually orthonormal, the matrix elements of the identity operator satisfy 
    \begin{eqnarray}
        \bra{\Psi'} \mathbb{I} \ket{\Psi} =& 1    & \text{, if } \Psi' = \Psi . 
    \end{eqnarray}
    Moreover, because the Bethe basis states are eigenstates of $H_{J}$ and $S^{z}$, their corresponding matrix elements are 
    \begin{eqnarray}
        \bra{\Psi'} H_{J} \ket{\Psi} =& E_{J}    & \text{, if } \Psi' = \Psi ,
        \\
        \bra{\Psi'} S^{z} \ket{\Psi} =& m        & \text{, if } \Psi' = \Psi .
        \label{eq: matrix element Sz}
    \end{eqnarray}
    Furthermore, the collective spin ladder operators $S^{+}$ and $S^{-}$ connect a Bethe basis state only to its immediate ancestor or descendant, respectively, i.e., 
    \begin{align}
        \bra{\Psi'} S^+ \ket{\Psi} &= \sqrt{s(s+1) - m(m+1)}, 
        \label{eq: matrix element S+} \\
            &\text{if $\Psi'$ is the immediate \;\ ancestor \;\ of $\Psi$}, \nonumber\\
        \bra{\Psi'} S^- \ket{\Psi} &= \sqrt{s(s+1) - m(m-1)}, 
        \label{eq: matrix element S-} \\
            &\text{if $\Psi'$ is the immediate descendant of $\Psi$}. \nonumber
    \end{align}
    Note that the matrix elements in Eq.~\eqref{eq: matrix element Sz}--\eqref{eq: matrix element S-} depend only on the quantum numbers $s$ and $m$, and are therefore identical for all multiplets with the same $\mathbb{S}$. Consequently, these operators share an identical block-matrix structure within every such multiplet, with only the diagonal matrix elements of $H_J$ generally differing between multiplets but taking a common value within each multiplet.
    
    For instance, in the Bethe basis $\mathcal{B}$, the matrix representation of the system Hamiltonian $H$ and the jump operators $L_{i}$ for $N=4$ are

    \begin{eqnarray}
        H &=&
        \begin{bmatrix}
            H^{\bf{0}, \, a} & 0 & 0 & 0 & 0 & 0 \\
            0 & H^{\bf{0}, \, b} & 0 & 0 & 0 & 0 \\
            0 & 0 & H^{\bf{1}, \, a} & 0 & 0 & 0 \\
            0 & 0 & 0 & H^{\bf{1}, \, b} & 0 & 0 \\
            0 & 0 & 0 & 0 & H^{\bf{1}, \, c} & 0 \\
            0 & 0 & 0 & 0 & 0 & H^{\bf{2}, \, a} \\
        \end{bmatrix}
        _\mathcal{B} ,
        \\
        L_{i} &=&
        \begin{bmatrix}
            L^{\bf{0}, \, a}_{i} & 0 & 0 & 0 & 0 & 0 \\
            0 & L^{\bf{0}, \, b}_{i} & 0 & 0 & 0 & 0 \\
            0 & 0 & L^{\bf{1}, \, a}_{i} & 0 & 0 & 0 \\
            0 & 0 & 0 & L^{\bf{1}, \, b}_{i} & 0 & 0 \\
            0 & 0 & 0 & 0 & L^{\bf{1}, \, c}_{i} & 0 \\
            0 & 0 & 0 & 0 & 0 & L^{\bf{2}, \, a}_{i} \\
        \end{bmatrix}
        _\mathcal{B} ,
    \end{eqnarray}
    where  
    \begin{eqnarray}
        H^{\bf{0}, \, a} &=&
        \begin{bmatrix}
            0 \\
        \end{bmatrix}
         ,
        \\
        H^{\bf{0}, \, b} &=&
        \begin{bmatrix}
            -2 J \\
        \end{bmatrix}
         ,
        \\
        H^{\bf{1}, \, a} &=&
        \begin{bmatrix}
            h^{z} & \sqrt{2} h^{-} & 0 \\
            \sqrt{2} h^{+} & 0 & \sqrt{2} h^{-} \\
            0 & \sqrt{2} h^{+} & -h^{z} \\
        \end{bmatrix}
         ,
        \\
        H^{\bf{1}, \, b} &=&
        \begin{bmatrix}
            - J + h^{z} & \sqrt{2} h^{-} & 0 \\
            \sqrt{2} h^{+} & - J & \sqrt{2} h^{-} \\
            0 & \sqrt{2} h^{+} & - J - h^{z} \\
        \end{bmatrix}
         ,
        \\
        H^{\bf{1}, \, c} &=&
        \begin{bmatrix}
            h^{z} & \sqrt{2} h^{-} & 0 \\
            \sqrt{2} h^{+} & 0 & \sqrt{2} h^{-} \\
            0 & \sqrt{2} h^{+} & -h^{z} \\
        \end{bmatrix}
         ,
        \\
        H^{\bf{2}, \, a} &=&
        \begin{bmatrix}
            J + 2 h^{z} & 2 h^{-} & 0 & 0 & 0 \\
            2 h^{+} & J + h^{z} & \sqrt{6} h^{-} & 0 & 0 \\
            0 & \sqrt{6} h^{+} & J & \sqrt{6} h^{-} & 0 \\
            0 & 0 & \sqrt{6} h^{+} & J-h^{z} & 2 h^{-} \\
            0 & 0 & 0 & 2 h^{+} & J-2 h^{z} \\
        \end{bmatrix}
    \end{eqnarray}      
        % \\
    \begin{eqnarray}
        L_{i}^{\bf{0}, \, a} &=&
        \begin{bmatrix}
            \ell^{0}_{i} \\
        \end{bmatrix}
        ,
        \\
        L_{i}^{\bf{0}, \, b} &=&
        \begin{bmatrix}
            \ell^{0}_{i} \\
        \end{bmatrix}
        ,
        \\
        L_{i}^{\bf{1}, \, a} &=&
        \begin{bmatrix}
            \ell^{0}_{i} + \ell^{z}_{i} & \sqrt{2} \ell^{-}_{i}  & 0 \\
            \sqrt{2} \ell^{+}_{i} & \ell^{0}_{i} & \sqrt{2} \ell^{-}_{i}  \\
            0 & \sqrt{2} \ell^{+}_{i} & \ell^{0}_{i} - \ell^{z}_{i} \\
        \end{bmatrix}
        ,
        \\
        L_{i}^{\bf{1}, \, b} &=&
        \begin{bmatrix}
            \ell^{0}_{i} + \ell^{z}_{i} & \sqrt{2} \ell^{-}_{i}  & 0 \\
            \sqrt{2} \ell^{+}_{i} & \ell^{0}_{i} & \sqrt{2} \ell^{-}_{i}  \\
            0 & \sqrt{2} \ell^{+}_{i} & \ell^{0}_{i} - \ell^{z}_{i} \\
        \end{bmatrix}
        ,
        \\
        L_{i}^{\bf{1}, \, c} &=&
        \begin{bmatrix}
            \ell^{0}_{i} + \ell^{z}_{i} & \sqrt{2} \ell^{-}_{i}  & 0 \\
            \sqrt{2} \ell^{+}_{i} & \ell^{0}_{i} & \sqrt{2} \ell^{-}_{i}  \\
            0 & \sqrt{2} \ell^{+}_{i} & \ell^{0}_{i} - \ell^{z}_{i} \\
        \end{bmatrix}
        ,
        \\
        L_{i}^{\bf{2}, \, a} &=&
        \begin{bmatrix}
            \ell^{0}_{i} + 2 \ell^{z}_{i} & 2 \ell^{-}_{i} & 0 & 0 & 0 \\
            2 \ell^{+}_{i} & \ell^{0}_{i} + \ell^{z}_{i} & \sqrt{6} \ell^{-}_{i} & 0 & 0 \\
            0 & \sqrt{6} \ell^{+}_{i} & \ell^{0}_{i} & \sqrt{6} \ell^{-}_{i} & 0 \\
            0 & 0 & \sqrt{6} \ell^{+}_{i} & \ell^{0}_{i} - \ell^{z}_{i} & 2 \ell^{-}_{i} \\
            0 & 0 & 0 & 2 \ell^{+}_{i} & \ell^{0}_{i} - 2 \ell^{z}_{i} \\
        \end{bmatrix} .
    \end{eqnarray}
    It is evident that the system Hamiltonian $H$ and the jump operators $L_i$ share the same block-diagonal structure, as each multiplet forms a block and each multiplet is not coupled with each other. Consequently, $6$ steady modes arise naturally, corresponding to the $6$ blocks.
    
    Moreover, the blocks of $L_i$ are identical across sectors with the same $\mathbb{S}$, namely $L_{i}^{\bf{0}, \, a} = L_{i}^{\bf{0}, \, b}$, and $L_{i}^{\bf{1}, \, a} = L_{i}^{\bf{1}, \, b} = L_{i}^{\bf{1}, \, c}$.
    In addition, the corresponding system Hamiltonian blocks differ only by multiples of identity, Consequently $8$ undamped modes emerge: $2$ of these arise from the fact that $H^{\bf{0}, \, a} - H^{\bf{0}, \, b} = 2J \, \mathbb{I}$, producing genuine persistent oscillatory modes with oscillatory frequencies $\omega = \pm 2J$, while $4$ more genuine persistent oscillatory modes result from $H^{\bf{1}, \, a} - H^{\bf{1}, \, b} = J \, \mathbb{I}$ and $H^{\bf{1}, \, c} - H^{\bf{1}, \, b} = J \, \mathbb{I}$, yielding oscillatory frequencies $\omega = \pm J$. The remaining $2$ are steady modes (regressed persistent oscillatory modes) due to $H^{\bf{1}, \, a} - H^{\bf{1}, \, c} = 0 \, \mathbb{I}$. Altogether, the system exhibits a total of $13$ undamped modes, of which $7$ steady modes and $6$ genuine persistent oscillatory modes.

\subsection{Characteristics of undamped modes for generic \textit{N}}

    The explicit expression of the Bethe basis and the associated matrix representations of $H$ and $L_{i}$ are unnecessary to characterize the undamped modes. The multiplicity $d_{\mathbb{S}}$ alone determines their number, while the $E_{J}$ of each multiplet set the oscillatory frequencies of the persistent oscillatory modes.
    
    For example, when $N=5$, the Bethe ansatz solution (Table~\ref{table: Bethe_basis_N_5}) shows the following:
    \begin{enumerate}
        \item There are $5$ multiplets with $\mathbb{S} = \frac{1}{2}$, and their corresponding $E_{J}$ are 
        $\left( \frac{-3 + 2\sqrt{5}}{4} J , \frac{-3 - 2\sqrt{5}}{4} J , \frac{-3}{4} J , \frac{-3 - 2\sqrt{5}}{4} J , \frac{-3 + 2\sqrt{5}}{4} J \right)$.
        These give $5^2 = 25$ undamped modes, of which:
        \begin{enumerate}
            \item $5$ are trivial steady modes,
            \item $4$ are regressed persistent oscillatory modes (steady modes, arising from  the two pairs of degenerate $E_{J}$), 
            \item $8$ are persistent oscillatory modes with frequency $\omega = \pm \left\lvert \frac{-3 + 2\sqrt{5}}{4} J - \frac{-3}{4} J \right\rvert = \pm \frac{\sqrt{5}}{2} J $,
            \item $8$ are persistent oscillatory modes with frequency $\omega = \pm \left\lvert \frac{-3 + 2\sqrt{5}}{4} J - \frac{-3 - 2\sqrt{5}}{4} J \right\rvert = \pm \sqrt{5} J $.
        \end{enumerate}
        \item There are $4$ multiplets with $\mathbb{S} = \frac{3}{2}$, and their corresponding $E_{J}$ are 
        $\left( \frac{\sqrt{5}}{4} J , \frac{- \sqrt{5}}{4} J , \frac{- \sqrt{5}}{4} J , \frac{\sqrt{5}}{4} J \right)$.
        These give $4^2 = 16$ undamped modes, of which:
        \begin{enumerate}
            \item $4$ are trivial steady modes,
            \item $4$ are regressed persistent oscillatory modes (steady modes, arising from  the two pairs of degenerate $E_{J}$), 
            \item $8$ are persistent oscillatory modes with frequency $\omega = \pm \left\lvert \frac{\sqrt{5}}{4} J - \frac{-\sqrt{5}}{4} J \right\rvert = \pm \frac{\sqrt{5}}{2} J $.
        \end{enumerate}
        \item Finally, the sole multiplet with $\mathbb{S} = \frac{5}{2}$, contributes to $1$ trivial steady mode.
    \end{enumerate}
    In summary, for $N=5$, the system exhibits a total of $42$ undamped modes, comprising $18$ steady modes, $16$ persistent oscillatory modes with frequency $\frac{\sqrt{5}}{2} J$, and $8$ persistent oscillatory modes with frequency $\sqrt{5} J$.
    
    For generic $N$, the number of undamped modes the system could support is
    \begin{eqnarray}
        \sum_{\mathbb{S} = \mathbb{S}_{min}}^{\mathbb{S}_{max}} (d_{\mathbb{S}})^2
        = C_{N} ,
    \end{eqnarray} 
    where $C_{N} = \frac{1}{N+1} \binom{2N}{N}$ denotes the $N$-th Catalan number.
    In the large $N$ limit, it has the asymptotic form $C_{N} \simeq \frac{4^N}{N^{3/2} \sqrt{\pi}}$, derived from Stirling’s approximation.
    Remarkably, these undamped modes are not just accidental, their existence is robust against perturbations in both the external magnetic field $\vec{h}$ and the specific form of the collective jump operators $L_i$.
    Their corresponding oscillatory frequencies are determined solely by the differences in $E_{J}$ between multiplets with the same $\mathbb{S}$.

    It is worth noting that extra undamped modes can also appear for specific choices of the jump operator parameters; these arise from the dark states~\cite{Buca_2012} of the jump operators. In the extreme case where $L_{i} = 0$, corresponding to a closed system, all $4^N$ Liouvillian eigenmodes become undamped. However, these modes are not the focus of this work, as they are fragile and highly sensitive to parameter variations.

\subsection{Nonlinear Collective Jump Operators}

    In the discussion above, we assumed for simplicity that the collective jump operators $L_{i}$ are linear. Nevertheless, the results, including the existence of undamped modes, their number, and oscillatory frequencies, remain valid even for nonlinear collective operators as in Eq.~\eqref{eq: nonlinear jump operator}. This is because the block-diagonal structure is preserved. In the nonlinear case, a collective jump operator can connect a Bethe basis state to all of its ancestor or descendant states, rather than only to the immediate ones. Crucially, states belonging to different multiplets remain disconnected, so the block-diagonal structure is maintained. The difference is that entries that were previously zero within the diagonal blocks of the jump operators may now be nonzero, yet these blocks remain identical across sectors with the same $\mathbb{S}$ because they share the same $s$ and $m$. Hence, the main results continue to hold.

\section{Conclusion and Outlook} 

    This study reveals the existence of undamped modes in a spin-1/2 Heisenberg chain coupled with an environment via collective spin jump operators. We identify the number of such modes for a generic system size $N$, with their (angular) oscillatory frequencies determined by gap of $E_{J}$, which can be derived from the Bethe ansatz solution. Remarkably, these modes are robust against variations in system parameters: both their number and oscillatory frequencies are insensitive to the specific form of the collective jump operators and the external field. This suggests that these modes are protected by the system’s structure and symmetries, rather than being fine-tuned artifacts. Such long-lived modes could serve as a valuable resource in quantum information processing, where stable, coherent excitations are essential for reliable operation.
    
    The approach developed in this work is expected to be applicable to a broad class of exactly solvable models via the Bethe ansatz. In addition to the Heisenberg XXX model considered in this work, future studies could extend this approach to models such as the Heisenberg XXZ model, Hubbard model, Gaudin-Yang model, and Lieb-Liniger model. While the detailed behavior and outcomes could differ depending on the structure and symmetries of each model, it is reasonable to anticipate that similar results might be obtained. This highlights the potential of the method to serve as a framework for studying coherent dynamics in a variety of open quantum, integrable systems, offering insights that could be broadly relevant across different physical contexts.
    
\section*{Acknowledgement}
    This work is supported by the EPSRC through Grant No. EP/W015641/1.

%=============================================================================

\bibliography{citation} % Produces the bibliography via BibTeX.

%=============================================================================

% --- Start appendix on a new page, one-column ---

\onecolumngrid
\clearpage
\appendix
\section{\label{bethe basis tables} Tables of Bethe Basis}

    In the tables below, $s$ denotes the collective spin quantum number, $m$ denotes the collective spin projection quantum number, $k_{i}$ denotes the quasi-momenta, and $E_{J}$ denotes the eigenenergy of the Heisenberg Hamiltonian $H_{J} \equiv \sum_{n=1}^{N} J \, \vec{S}_{n} \cdot \vec{S}_{n+1}$ with coupling constant $J$.

    Basis states are arranged into multiplets, with the state at the top of each sector representing the highest-weight state and the states below serving as its associated descendants.

    {\renewcommand{\arraystretch}{1.25} % Increase row height
        \begin{table}[!h]
        \centering
            \caption{Bethe basis $\mathcal{B}$ for the (one-dimensional, periodic boundary) Heisenberg XXX model with $N=3$ qubits.}
            \label{table: Bethe_basis_N_3}
                
        \begin{tabular}{||cc|ccc|c|c||}
            \hline 
                $s$   & $m$    & $k_{1}$     & $k_{2}$ & $k_{3}$ & $E_{J} / J$ & Normalized Bethe Basis States $\ket{ \Psi \{ k_{\mu} \} }$ 
            \\ \hline \hline 
                $1/2$ & $1/2$  & $2 \pi / 3$ & --      & --      & $-3/4$      & 
                $\frac{1}{\sqrt{3}} \left( + e^{\mathbbm{i} 2 \pi /3} \ket{ \downarrow\uparrow\uparrow } + e^{\mathbbm{i} 4 \pi /3} \ket{ \uparrow\downarrow\uparrow } + \phantom{} \ket{ \uparrow\uparrow\downarrow } \right)$ 
            \\ 
                $1/2$ & $-1/2$ & $2 \pi / 3$ & $0$     & --      & $-3/4$      & 
                $\frac{1}{\sqrt{3}} \left( + e^{\mathbbm{i} 4 \pi /3} \ket{ \uparrow\downarrow\downarrow } + e^{\mathbbm{i} 2 \pi /3} \ket{ \downarrow\uparrow\downarrow } + \phantom{} \ket{ \downarrow\downarrow\uparrow } \right)$ 
            \\ \hline 
                $1/2$ & $1/2$  & $4 \pi / 3$ & --      & --      & $-3/4$      & 
                $\frac{1}{\sqrt{3}} \left( + e^{\mathbbm{i} 4 \pi /3} \ket{ \downarrow\uparrow\uparrow } + e^{\mathbbm{i} 2 \pi /3} \ket{ \uparrow\downarrow\uparrow } + \phantom{} \ket{ \uparrow\uparrow\downarrow } \right)$ 
            \\ 
                $1/2$ & $-1/2$ & $4 \pi / 3$ & $0$     & --      & $-3/4$      & 
                $\frac{1}{\sqrt{3}} \left( + e^{\mathbbm{i} 2 \pi /3} \ket{ \uparrow\downarrow\downarrow } + e^{\mathbbm{i} 4 \pi /3} \ket{ \downarrow\uparrow\downarrow } + \phantom{} \ket{ \downarrow\downarrow\uparrow } \right)$ 
            \\ \hline 
                $3/2$ & $3/2$  & --          & --      & --      & $3/4$       & 
                $\phantom{+\frac{1}{\sqrt{1}}}$ $\ket{ \uparrow\uparrow\uparrow }$
            \\ 
                $3/2$ & $1/2$  & $0$         & --      & --      & $3/4$       & 
                $\frac{1}{\sqrt{3}} \left( + \ket{ \downarrow\uparrow\uparrow } + \ket{ \uparrow\downarrow\uparrow } + \ket{ \uparrow\uparrow\downarrow } \right)$ 
            \\ 
                $3/2$ & $-1/2$ & $0$         & $0$     & --      & $3/4$       & 
                $\frac{1}{\sqrt{3}} \left( + \ket{ \uparrow\downarrow\downarrow } + \ket{ \downarrow\uparrow\downarrow } + \ket{ \downarrow\downarrow\uparrow } \right)$ 
            \\ 
                $3/2$ & $-3/2$ & $0$         & $0$     & $0$     & $3/4$       & 
                $\phantom{+\frac{1}{\sqrt{1}}}$ $\ket{ \downarrow\downarrow\downarrow }$ 
            \\ \hline 
        \end{tabular}
        
        \end{table}
        \begin{table}[!h]
        \centering
            \caption{Bethe basis $\mathcal{B}$ for the (one-dimensional, periodic boundary) Heisenberg XXX model with $N=4$ qubits. 
            \newline Adapted from~\cite{10.1063/1.4822511}.}
            \label{table: Bethe_basis_N_4}
                
    \footnotesize
        \begin{tabular}{||cc|cccc|c|c||}
            \hline 
                $s$ & $m$  & $k_{1}$                        & $k_{2}$                        & $k_{3}$ & $k_{4}$ & $E_{J} / J$ & Normalized Bethe Basis States $\ket{ \Psi \{ k_{\mu} \} }$ 
            \\ \hline \hline 
                $0$ & $0$  & $\pi / 2 + \mathbbm{i} \infty$ & $\pi / 2 - \mathbbm{i} \infty$ & --      & --      & $0$         & 
                $\frac{1}{\sqrt{4}} \left( + \ket{ \downarrow\downarrow\uparrow\uparrow } - \ket{ \uparrow\downarrow\downarrow\uparrow } + \ket{ \uparrow\uparrow\downarrow\downarrow } - \ket{ \downarrow\uparrow\uparrow\downarrow } \right)$ 
            \\ \hline 
                $0$ & $0$  & $2 \pi / 3$                    & $4 \pi / 3$                    & --      & --      & $-2$        & 
                $\frac{1}{\sqrt{12}} \left( + \ket{ \downarrow\downarrow\uparrow\uparrow } + \ket{ \uparrow\downarrow\downarrow\uparrow } + \ket{ \uparrow\uparrow\downarrow\downarrow } + \ket{ \downarrow\uparrow\uparrow\downarrow } 
                -2 \ket{ \downarrow\uparrow\downarrow\uparrow } -2 \ket{ \uparrow\downarrow\uparrow\downarrow } \right)$ 
            \\ \hline 
                $1$ & $1$  & $\pi / 2$                      & --                             & --      & --      & $0$         & 
                $\frac{1}{\sqrt{4}} \left( +\mathbbm{i} \ket{ \downarrow\uparrow\uparrow\uparrow } - \ket{ \uparrow\downarrow\uparrow\uparrow } -\mathbbm{i} \ket{ \uparrow\uparrow\downarrow\uparrow } + \ket{ \uparrow\uparrow\uparrow\downarrow } \right)$ 
            \\ 
                $1$ & $0$  & $\pi / 2$                      & $0$                            & --      & --      & $0$         & 
                $\frac{1}{\sqrt{4}} \left( e^{\mathbbm{i} 3 \pi /4} \ket{ \downarrow\downarrow\uparrow\uparrow } + e^{\mathbbm{i} 5 \pi /4} \ket{ \uparrow\downarrow\downarrow\uparrow } + e^{\mathbbm{i} 7 \pi /4} \ket{ \uparrow\uparrow\downarrow\downarrow } + e^{\mathbbm{i} \pi /4} \ket{ \downarrow\uparrow\uparrow\downarrow } \right)$ 
            \\ 
                $1$ & $-1$ & $\pi / 2$                      & $0$                            & $0$     & --      & $0$         & 
                $\frac{1}{\sqrt{4}} \left( -\mathbbm{i} \ket{ \uparrow\downarrow\downarrow\downarrow } + \ket{ \downarrow\uparrow\downarrow\downarrow } +\mathbbm{i} \ket{ \downarrow\downarrow\uparrow\downarrow } - \ket{ \downarrow\downarrow\downarrow\uparrow } \right)$ 
            \\ \hline 
                $1$ & $1$  & $\pi$                          & --                             & --      & --      & $-1$        & 
                $\frac{1}{\sqrt{4}} \left( - \ket{ \downarrow\uparrow\uparrow\uparrow } + \ket{ \uparrow\downarrow\uparrow\uparrow } - \ket{ \uparrow\uparrow\downarrow\uparrow } + \ket{ \uparrow\uparrow\uparrow\downarrow } \right)$  
            \\ 
                $1$ & $0$  & $\pi$                          & $0$                            & --      & --      & $-1$        & 
                $\frac{1}{\sqrt{2}} \left( - \ket{ \downarrow\uparrow\downarrow\uparrow } + \ket{ \uparrow\downarrow\uparrow\downarrow } \right)$ 
            \\ 
                $1$ & $-1$ & $\pi$                          & $0$                            & $0$     & --      & $-1$        & 
                $\frac{1}{\sqrt{4}} \left( + \ket{ \uparrow\downarrow\downarrow\downarrow } - \ket{ \downarrow\uparrow\downarrow\downarrow } + \ket{ \downarrow\downarrow\uparrow\downarrow } - \ket{ \downarrow\downarrow\downarrow\uparrow } \right)$ 
            \\ \hline 
                $1$ & $1$  & $3 \pi / 2$                    & --                             & --      & --      & $0$         & 
                $\frac{1}{\sqrt{4}} \left( -\mathbbm{i} \ket{ \downarrow\uparrow\uparrow\uparrow } - \ket{ \uparrow\downarrow\uparrow\uparrow } +\mathbbm{i} \ket{ \uparrow\uparrow\downarrow\uparrow } + \ket{ \uparrow\uparrow\uparrow\downarrow } \right)$ 
            \\ 
                $1$ & $0$  & $3 \pi / 2$                    & $0$                            & --      & --      & $0$         & 
                $\frac{1}{\sqrt{4}} \left( e^{\mathbbm{i} 5 \pi /4} \ket{ \downarrow\downarrow\uparrow\uparrow } + e^{\mathbbm{i} 3 \pi /4} \ket{ \uparrow\downarrow\downarrow\uparrow } + e^{\mathbbm{i} \pi /4} \ket{ \uparrow\uparrow\downarrow\downarrow } + e^{\mathbbm{i} 7 \pi /4} \ket{ \downarrow\uparrow\uparrow\downarrow } \right)$ 
            \\ 
                $1$ & $-1$ & $3 \pi / 2$                    & $0$                            & $0$     & --      & $0$         & 
                $\frac{1}{\sqrt{4}} \left( +\mathbbm{i} \ket{ \uparrow\downarrow\downarrow\downarrow } + \ket{ \downarrow\uparrow\downarrow\downarrow } -\mathbbm{i} \ket{ \downarrow\downarrow\uparrow\downarrow } - \ket{ \downarrow\downarrow\downarrow\uparrow } \right)$ 
            \\ \hline 
                $2$ & $2$  & --                             & --                             & --      & --      & $1$         & 
                $\ket{ \uparrow\uparrow\uparrow\uparrow }$
                $\phantom{\frac{}{\sqrt{}}}$
            \\ 
                $2$ & $1$  & $0$                            & --                             & --      & --      & $1$         & 
                $\frac{1}{\sqrt{4}} \left( + \ket{ \downarrow\uparrow\uparrow\uparrow } + \ket{ \uparrow\downarrow\uparrow\uparrow } + \ket{ \uparrow\uparrow\downarrow\uparrow } + \ket{ \uparrow\uparrow\uparrow\downarrow } \right)$ 
            \\ 
                $2$ & $0$  & $0$                            & $0$                            & --      & --      & $1$         & 
                $\frac{1}{\sqrt{6}} \left( + \ket{ \downarrow\downarrow\uparrow\uparrow } + \ket{ \uparrow\downarrow\downarrow\uparrow } + \ket{ \uparrow\uparrow\downarrow\downarrow } + \ket{ \downarrow\uparrow\uparrow\downarrow } 
                + \ket{ \downarrow\uparrow\downarrow\uparrow } + \ket{ \uparrow\downarrow\uparrow\downarrow } \right)$ 
            \\ 
                $2$ & $-1$ & $0$                            & $0$                            & $0$     & --      & $1$         & 
                $\frac{1}{\sqrt{4}} \left( + \ket{ \uparrow\downarrow\downarrow\downarrow } + \ket{ \downarrow\uparrow\downarrow\downarrow } + \ket{ \downarrow\downarrow\uparrow\downarrow } + \ket{ \downarrow\downarrow\downarrow\uparrow } \right)$ 
            \\ 
                $2$ & $-2$ & $0$                            & $0$                            & $0$     & $0$     & $1$         & 
                $\ket{ \downarrow\downarrow\downarrow\downarrow }$ $\phantom{\frac{}{\sqrt{}}}$
            \\ \hline 
        \end{tabular}
        
        \end{table}
        \begin{table}[!p]
        \centering
            \caption{Bethe basis $\mathcal{B}$ for the (one-dimensional, periodic boundary) Heisenberg XXX model with $N=5$ qubits. 
            \newline Adapted from~\cite{10.1063/1.4822511}.}
            \label{table: Bethe_basis_N_5}
                    
        \scriptsize
        \begin{tabular}{||cc|ccccc|c|c||}
            \hline 
                $s$   & $m$    & $k_{1}$                          & $k_{2}$                          & $k_{3}$ & $k_{4}$ & $k_{5}$ & $E_{J} / J$                                & 
                Normalized Bethe Basis States $\ket{ \Psi \{ k_{\mu} \} }$ 
            \\ \hline \hline 
                $1/2$ & $1/2$  & $2 \pi / 5 + \mathbbm{i} 1.1989$ & $2 \pi / 5 - \mathbbm{i} 1.1989$ & --      & --      & --      & $\frac{-3 + 2\sqrt{5}}{4} \approx 0.3680$  & 
                $\frac{1}{\sqrt{15}} 
                \begin{pmatrix}
                    \phantom{+} \frac{1 + \sqrt{5}}{2} \left( + e^{\mathbbm{i} 6 \pi /5} \ket{ \downarrow\downarrow\uparrow\uparrow\uparrow } + \phantom{e^{\mathbbm{i} 0 \pi /5}} \ket{ \uparrow\downarrow\downarrow\uparrow\uparrow } + e^{\mathbbm{i} 4 \pi /5} \ket{ \uparrow\uparrow\downarrow\downarrow\uparrow } + e^{\mathbbm{i} 8 \pi /5} \ket{ \uparrow\uparrow\uparrow\downarrow\downarrow } + e^{\mathbbm{i} 2 \pi /5} \ket{ \downarrow\uparrow\uparrow\uparrow\downarrow } \right) 
                    \\ 
                    + \frac{2}{1 + \sqrt{5}} \left( + e^{\mathbbm{i} 6 \pi /5} \ket{ \uparrow\uparrow\downarrow\uparrow\downarrow } + \phantom{e^{\mathbbm{i} 0 \pi /5}} \ket{ \downarrow\uparrow\uparrow\downarrow\uparrow } + e^{\mathbbm{i} 4 \pi /5} \ket{ \uparrow\downarrow\uparrow\uparrow\downarrow } + e^{\mathbbm{i} 8 \pi /5} \ket{ \downarrow\uparrow\downarrow\uparrow\uparrow } + e^{\mathbbm{i} 2 \pi /5} \ket{ \uparrow\downarrow\uparrow\downarrow\uparrow } \right) 
                    \end{pmatrix}$ 
            \\ 
                $1/2$ & $-1/2$ & $2 \pi / 5 + \mathbbm{i} 1.1989$ & $2 \pi / 5 - \mathbbm{i} 1.1989$ & $0$     & --      & --      & $\frac{-3 + 2\sqrt{5}}{4} \approx 0.3680$  & 
                $\frac{1}{\sqrt{15}} 
                \begin{pmatrix}
                    \phantom{+} \frac{1 + \sqrt{5}}{2} \left( + e^{\mathbbm{i} 6 \pi /5} \ket{ \uparrow\uparrow\downarrow\downarrow\downarrow } + \phantom{e^{\mathbbm{i} 0 \pi /5}} \ket{ \downarrow\uparrow\uparrow\downarrow\downarrow } + e^{\mathbbm{i} 4 \pi /5} \ket{ \downarrow\downarrow\uparrow\uparrow\downarrow } + e^{\mathbbm{i} 8 \pi /5} \ket{ \downarrow\downarrow\downarrow\uparrow\uparrow } + e^{\mathbbm{i} 2 \pi /5} \ket{ \uparrow\downarrow\downarrow\downarrow\uparrow } \right) 
                    \\ 
                    + \frac{2}{1 + \sqrt{5}} \left( + e^{\mathbbm{i} 6 \pi /5} \ket{ \downarrow\downarrow\uparrow\downarrow\uparrow } + \phantom{e^{\mathbbm{i} 0 \pi /5}} \ket{ \uparrow\downarrow\downarrow\uparrow\downarrow } + e^{\mathbbm{i} 4 \pi /5} \ket{ \downarrow\uparrow\downarrow\downarrow\uparrow } + e^{\mathbbm{i} 8 \pi /5} \ket{ \uparrow\downarrow\uparrow\downarrow\downarrow } + e^{\mathbbm{i} 2 \pi /5} \ket{ \downarrow\uparrow\downarrow\uparrow\downarrow } \right) 
                    \end{pmatrix}$ 
            \\ \hline 
                $1/2$ & $1/2$  & $1.7053$                         & $3.3212$                         & --      & --      & --      & $\frac{-3 - 2\sqrt{5}}{4} \approx -1.8680$ & 
                $\frac{1}{\sqrt{15}} 
                \begin{pmatrix}
                    \phantom{+} \frac{2}{1 + \sqrt{5}} \left( + e^{\mathbbm{i} 7 \pi /5} \ket{ \downarrow\downarrow\uparrow\uparrow\uparrow } + e^{\mathbbm{i} 5 \pi /5} \ket{ \uparrow\downarrow\downarrow\uparrow\uparrow } + e^{\mathbbm{i} 3 \pi /5} \ket{ \uparrow\uparrow\downarrow\downarrow\uparrow } + e^{\mathbbm{i} \pi /5 \phantom{1}} \ket{ \uparrow\uparrow\uparrow\downarrow\downarrow } + e^{\mathbbm{i} 9 \pi /5} \ket{ \downarrow\uparrow\uparrow\uparrow\downarrow } \right) 
                    \\ 
                    + \frac{1 + \sqrt{5}}{2} \left( + e^{\mathbbm{i} 7 \pi /5} \ket{ \uparrow\uparrow\downarrow\uparrow\downarrow } + e^{\mathbbm{i} 5 \pi /5} \ket{ \downarrow\uparrow\uparrow\downarrow\uparrow } + e^{\mathbbm{i} 3 \pi /5} \ket{ \uparrow\downarrow\uparrow\uparrow\downarrow } + e^{\mathbbm{i} \pi /5 \phantom{1}} \ket{ \downarrow\uparrow\downarrow\uparrow\uparrow } + e^{\mathbbm{i} 9 \pi /5} \ket{ \uparrow\downarrow\uparrow\downarrow\uparrow } \right) 
                    \end{pmatrix}$ 
            \\ 
                $1/2$ & $-1/2$ & $1.7053$                         & $3.3212$                         & $0$     & --      & --      & $\frac{-3 - 2\sqrt{5}}{4} \approx -1.8680$ & 
                $\frac{1}{\sqrt{15}} 
                \begin{pmatrix}
                    \phantom{+} \frac{2}{1 + \sqrt{5}} \left( + e^{\mathbbm{i} 7 \pi /5} \ket{ \uparrow\uparrow\downarrow\downarrow\downarrow } + e^{\mathbbm{i} 5 \pi /5} \ket{ \downarrow\uparrow\uparrow\downarrow\downarrow } + e^{\mathbbm{i} 3 \pi /5} \ket{ \downarrow\downarrow\uparrow\uparrow\downarrow } + e^{\mathbbm{i} \pi /5 \phantom{1}} \ket{ \downarrow\downarrow\downarrow\uparrow\uparrow } + e^{\mathbbm{i} 9 \pi /5} \ket{ \uparrow\downarrow\downarrow\downarrow\uparrow } \right) 
                    \\ 
                    + \frac{1 + \sqrt{5}}{2} \left( + e^{\mathbbm{i} 7 \pi /5} \ket{ \downarrow\downarrow\uparrow\downarrow\uparrow } + e^{\mathbbm{i} 5 \pi /5} \ket{ \uparrow\downarrow\downarrow\uparrow\downarrow } + e^{\mathbbm{i} 3 \pi /5} \ket{ \downarrow\uparrow\downarrow\downarrow\uparrow } + e^{\mathbbm{i} \pi /5 \phantom{1}} \ket{ \uparrow\downarrow\uparrow\downarrow\downarrow } + e^{\mathbbm{i} 9 \pi /5} \ket{ \downarrow\uparrow\downarrow\uparrow\downarrow } \right) 
                    \end{pmatrix}$ 
            \\ \hline 
                $1/2$ & $1/2$  & $\pi / 2$                        & $3 \pi / 2$                      & --      & --      & --      & $-\frac{3}{4} = -0.75$                     & 
                $\frac{1}{\sqrt{10}} 
                \begin{pmatrix}
                    + \ket{ \downarrow\downarrow\uparrow\uparrow\uparrow } + \ket{ \uparrow\downarrow\downarrow\uparrow\uparrow } + \ket{ \uparrow\uparrow\downarrow\downarrow\uparrow } + \ket{ \uparrow\uparrow\uparrow\downarrow\downarrow } + \ket{ \downarrow\uparrow\uparrow\uparrow\downarrow } 
                    \\
                    - \ket{ \uparrow\uparrow\downarrow\uparrow\downarrow } - \ket{ \downarrow\uparrow\uparrow\downarrow\uparrow } - \ket{ \uparrow\downarrow\uparrow\uparrow\downarrow } - \ket{ \downarrow\uparrow\downarrow\uparrow\uparrow } - \ket{ \uparrow\downarrow\uparrow\downarrow\uparrow } 
                \end{pmatrix}$ 
            \\ 
                $1/2$ & $-1/2$ & $\pi / 2$                        & $3 \pi / 2$                      & $0$     & --      & --      & $-\frac{3}{4} = -0.75$                     & 
                $\frac{1}{\sqrt{10}} 
                \begin{pmatrix} 
                    + \ket{ \uparrow\uparrow\downarrow\downarrow\downarrow } + \ket{ \downarrow\uparrow\uparrow\downarrow\downarrow } + \ket{ \downarrow\downarrow\uparrow\uparrow\downarrow } + \ket{ \downarrow\downarrow\downarrow\uparrow\uparrow } + \ket{ \uparrow\downarrow\downarrow\downarrow\uparrow } 
                    \\
                    - \ket{ \downarrow\downarrow\uparrow\downarrow\uparrow } - \ket{ \uparrow\downarrow\downarrow\uparrow\downarrow } - \ket{ \downarrow\uparrow\downarrow\downarrow\uparrow } - \ket{ \uparrow\downarrow\uparrow\downarrow\downarrow } - \ket{ \downarrow\uparrow\downarrow\uparrow\downarrow } 
                \end{pmatrix}$ 
            \\ \hline 
                $1/2$ & $1/2$  & $2.9620$                         & $4.5779$                         & --      & --      & --      & $\frac{-3 - 2\sqrt{5}}{4} \approx -1.8680$ & 
                $\frac{1}{\sqrt{15}} 
                \begin{pmatrix}
                    \phantom{+} \frac{2}{1 + \sqrt{5}} \left( + e^{\mathbbm{i} 3 \pi /5} \ket{ \downarrow\downarrow\uparrow\uparrow\uparrow } + e^{\mathbbm{i} 5 \pi /5} \ket{ \uparrow\downarrow\downarrow\uparrow\uparrow } + e^{\mathbbm{i} 7 \pi /5} \ket{ \uparrow\uparrow\downarrow\downarrow\uparrow } + e^{\mathbbm{i} 9 \pi /5} \ket{ \uparrow\uparrow\uparrow\downarrow\downarrow } + e^{\mathbbm{i} \pi /5 \phantom{1}} \ket{ \downarrow\uparrow\uparrow\uparrow\downarrow } \right) 
                    \\ 
                    + \frac{1 + \sqrt{5}}{2} \left( + e^{\mathbbm{i} 3 \pi /5} \ket{ \uparrow\uparrow\downarrow\uparrow\downarrow } + e^{\mathbbm{i} 5 \pi /5} \ket{ \downarrow\uparrow\uparrow\downarrow\uparrow } + e^{\mathbbm{i} 7 \pi /5} \ket{ \uparrow\downarrow\uparrow\uparrow\downarrow } + e^{\mathbbm{i} 9 \pi /5} \ket{ \downarrow\uparrow\downarrow\uparrow\uparrow } + e^{\mathbbm{i} \pi /5 \phantom{1}} \ket{ \uparrow\downarrow\uparrow\downarrow\uparrow } \right) 
                    \end{pmatrix}$ 
            \\ 
                $1/2$ & $-1/2$ & $2.9620$                         & $4.5779$                         & $0$     & --      & --      & $\frac{-3 - 2\sqrt{5}}{4} \approx -1.8680$ & 
                $\frac{1}{\sqrt{15}} 
                \begin{pmatrix}
                    \phantom{+} \frac{2}{1 + \sqrt{5}} \left( + e^{\mathbbm{i} 3 \pi /5} \ket{ \uparrow\uparrow\downarrow\downarrow\downarrow } + e^{\mathbbm{i} 5\pi /5} \ket{ \downarrow\uparrow\uparrow\downarrow\downarrow } + e^{\mathbbm{i} 7 \pi /5} \ket{ \downarrow\downarrow\uparrow\uparrow\downarrow } + e^{\mathbbm{i} 9 \pi /5} \ket{ \downarrow\downarrow\downarrow\uparrow\uparrow } + e^{\mathbbm{i} \pi /5 \phantom{1}} \ket{ \uparrow\downarrow\downarrow\downarrow\uparrow } \right) 
                    \\ 
                    + \frac{1 + \sqrt{5}}{2} \left( + e^{\mathbbm{i} 3 \pi /5} \ket{ \downarrow\downarrow\uparrow\downarrow\uparrow } + e^{\mathbbm{i} 5 \pi /5} \ket{ \uparrow\downarrow\downarrow\uparrow\downarrow } + e^{\mathbbm{i} 7 \pi /5} \ket{ \downarrow\uparrow\downarrow\downarrow\uparrow } + e^{\mathbbm{i} 9 \pi /5} \ket{ \uparrow\downarrow\uparrow\downarrow\downarrow } + e^{\mathbbm{i} \pi /5 \phantom{1}} \ket{ \downarrow\uparrow\downarrow\uparrow\downarrow } \right) 
                    \end{pmatrix}$ 
            \\ \hline 
                $1/2$ & $1/2$  & $8 \pi / 5 + \mathbbm{i} 1.1989$ & $8 \pi / 5 - \mathbbm{i} 1.1989$ & --      & --      & --      & $\frac{-3 + 2\sqrt{5}}{4} \approx 0.3680$  & 
                $\frac{1}{\sqrt{15}} 
                \begin{pmatrix}
                    \phantom{+} \frac{1 + \sqrt{5}}{2} \left( + e^{\mathbbm{i} 4 \pi /5} \ket{ \downarrow\downarrow\uparrow\uparrow\uparrow } + \phantom{e^{\mathbbm{i} 0 \pi /5}} \ket{ \uparrow\downarrow\downarrow\uparrow\uparrow } + e^{\mathbbm{i} 6 \pi /5} \ket{ \uparrow\uparrow\downarrow\downarrow\uparrow } + e^{\mathbbm{i} 2 \pi /5} \ket{ \uparrow\uparrow\uparrow\downarrow\downarrow } + e^{\mathbbm{i} 8 \pi /5} \ket{ \downarrow\uparrow\uparrow\uparrow\downarrow } \right) 
                    \\ 
                    + \frac{2}{1 + \sqrt{5}} \left( + e^{\mathbbm{i} 4 \pi /5} \ket{ \uparrow\uparrow\downarrow\uparrow\downarrow } + \phantom{e^{\mathbbm{i} 0 \pi /5}} \ket{ \downarrow\uparrow\uparrow\downarrow\uparrow } + e^{\mathbbm{i} 6 \pi /5} \ket{ \uparrow\downarrow\uparrow\uparrow\downarrow } + e^{\mathbbm{i} 2 \pi /5} \ket{ \downarrow\uparrow\downarrow\uparrow\uparrow } + e^{\mathbbm{i} 8 \pi /5} \ket{ \uparrow\downarrow\uparrow\downarrow\uparrow } \right) 
                    \end{pmatrix}$ 
            \\ 
                $1/2$ & $-1/2$ & $8 \pi / 5 + \mathbbm{i} 1.1989$ & $8 \pi / 5 - \mathbbm{i} 1.1989$ & $0$     & --      & --      & $\frac{-3 + 2\sqrt{5}}{4} \approx 0.3680$  & 
                $\frac{1}{\sqrt{15}} 
                \begin{pmatrix}
                    \phantom{+} \frac{1 + \sqrt{5}}{2} \left( + e^{\mathbbm{i} 4 \pi /5} \ket{ \uparrow\uparrow\downarrow\downarrow\downarrow } + \phantom{e^{\mathbbm{i} 0 \pi /5}} \ket{ \downarrow\uparrow\uparrow\downarrow\downarrow } + e^{\mathbbm{i} 6 \pi /5} \ket{ \downarrow\downarrow\uparrow\uparrow\downarrow } + e^{\mathbbm{i} 2 \pi /5} \ket{ \downarrow\downarrow\downarrow\uparrow\uparrow } + e^{\mathbbm{i} 8 \pi /5} \ket{ \uparrow\downarrow\downarrow\downarrow\uparrow } \right) 
                    \\ 
                    + \frac{2}{1 + \sqrt{5}} \left( + e^{\mathbbm{i} 4 \pi /5} \ket{ \downarrow\downarrow\uparrow\downarrow\uparrow } + \phantom{e^{\mathbbm{i} 0 \pi /5}} \ket{ \uparrow\downarrow\downarrow\uparrow\downarrow } + e^{\mathbbm{i} 6 \pi /5} \ket{ \downarrow\uparrow\downarrow\downarrow\uparrow } + e^{\mathbbm{i} 2 \pi /5} \ket{ \uparrow\downarrow\uparrow\downarrow\downarrow } + e^{\mathbbm{i} 8 \pi /5} \ket{ \downarrow\uparrow\downarrow\uparrow\downarrow } \right) 
                    \end{pmatrix}$ 
            \\ \hline 
                $3/2$ & $3/2$  & $2 \pi / 5$                      & --                               & --      & --      & --      & $\frac{\sqrt{5}}{4} \approx 0.5590$        & 
                $\frac{1}{\sqrt{5}} \left( + e^{\mathbbm{i} 2 \pi /5} \ket{ \downarrow\uparrow\uparrow\uparrow\uparrow } + e^{\mathbbm{i} 4 \pi /5} \ket{ \uparrow\downarrow\uparrow\uparrow\uparrow } + e^{\mathbbm{i} 6 \pi /5} \ket{ \uparrow\uparrow\downarrow\uparrow\uparrow } + e^{\mathbbm{i} 8 \pi /5} \ket{ \uparrow\uparrow\uparrow\downarrow\uparrow } + \phantom{e^{\mathbbm{i} 0 \pi /5}} \ket{ \uparrow\uparrow\uparrow\uparrow\downarrow } \right)$ 
            \\
                $3/2$ & $1/2$  & $2 \pi / 5$                      & $0$                              & --      & --      & --      & $\frac{\sqrt{5}}{4} \approx 0.5590$        & 
                $\frac{1}{\sqrt{15}} 
                \begin{pmatrix}
                    \phantom{+} \frac{1 + \sqrt{5}}{2} \left( + e^{\mathbbm{i} 3 \pi /5} \ket{ \downarrow\downarrow\uparrow\uparrow\uparrow } + e^{\mathbbm{i} 5 \pi /5} \ket{ \uparrow\downarrow\downarrow\uparrow\uparrow } + e^{\mathbbm{i} 7 \pi /5} \ket{ \uparrow\uparrow\downarrow\downarrow\uparrow } + e^{\mathbbm{i} 9 \pi /5} \ket{ \uparrow\uparrow\uparrow\downarrow\downarrow } + e^{\mathbbm{i} \pi /5 \phantom{1}} \ket{ \downarrow\uparrow\uparrow\uparrow\downarrow } \right) 
                    \\ 
                    + \frac{2}{1 + \sqrt{5}} \left( + e^{\mathbbm{i} 8 \pi /5} \ket{ \uparrow\uparrow\downarrow\uparrow\downarrow } + \phantom{e^{\mathbbm{i} 0 \pi /5}} \ket{ \downarrow\uparrow\uparrow\downarrow\uparrow } + e^{\mathbbm{i} 2 \pi /5} \ket{ \uparrow\downarrow\uparrow\uparrow\downarrow } + e^{\mathbbm{i} 4 \pi /5} \ket{ \downarrow\uparrow\downarrow\uparrow\uparrow } + e^{\mathbbm{i} 6 \pi /5} \ket{ \uparrow\downarrow\uparrow\downarrow\uparrow } \right) 
                    \end{pmatrix}$ 
            \\ 
                $3/2$ & $-1/2$ & $2 \pi / 5$                      & $0$                              & $0$     & --      & --      & $\frac{\sqrt{5}}{4} \approx 0.5590$        & 
                $\frac{1}{\sqrt{15}} 
                \begin{pmatrix}
                    \phantom{+} \frac{1 + \sqrt{5}}{2} \left( + e^{\mathbbm{i} 8 \pi /5} \ket{ \uparrow\uparrow\downarrow\downarrow\downarrow } + \phantom{e^{\mathbbm{i} 0 \pi /5}} \ket{ \downarrow\uparrow\uparrow\downarrow\downarrow } + e^{\mathbbm{i} 2 \pi /5} \ket{ \downarrow\downarrow\uparrow\uparrow\downarrow } + e^{\mathbbm{i} 4 \pi /5} \ket{ \downarrow\downarrow\downarrow\uparrow\uparrow } + e^{\mathbbm{i} 6 \pi /5} \ket{ \uparrow\downarrow\downarrow\downarrow\uparrow } \right) 
                    \\ 
                    + \frac{2}{1 + \sqrt{5}} \left( + e^{\mathbbm{i} 3 \pi /5} \ket{ \downarrow\downarrow\uparrow\downarrow\uparrow } + e^{\mathbbm{i} 5 \pi /5} \ket{ \uparrow\downarrow\downarrow\uparrow\downarrow } + e^{\mathbbm{i} 7 \pi /5} \ket{ \downarrow\uparrow\downarrow\downarrow\uparrow } + e^{\mathbbm{i} 9 \pi /5} \ket{ \uparrow\downarrow\uparrow\downarrow\downarrow } + e^{\mathbbm{i} \pi /5 \phantom{1}} \ket{ \downarrow\uparrow\downarrow\uparrow\downarrow } \right) 
                    \end{pmatrix}$ 
            \\ 
                $3/2$ & $-3/2$ & $2 \pi / 5$                      & $0$                              & $0$     & $0$     & --      & $\frac{\sqrt{5}}{4} \approx 0.5590$        & 
                $\frac{1}{\sqrt{5}} \left( + e^{\mathbbm{i} 7 \pi /5} \ket{ \uparrow\downarrow\downarrow\downarrow\downarrow } + e^{\mathbbm{i} 9 \pi /5} \ket{ \downarrow\uparrow\downarrow\downarrow\downarrow } + e^{\mathbbm{i} \pi /5 \phantom{1}} \ket{ \downarrow\downarrow\uparrow\downarrow\downarrow } + e^{\mathbbm{i} 3 \pi /5} \ket{ \downarrow\downarrow\downarrow\uparrow\downarrow } + e^{\mathbbm{i} 5 \pi /5} \ket{ \downarrow\downarrow\downarrow\downarrow\uparrow } \right)$ 
            \\ \hline 
                $3/2$ & $3/2$  & $4 \pi / 5$                      & --                               & --      & --      & --      & $-\frac{\sqrt{5}}{4} \approx -0.5590$      & 
                $\frac{1}{\sqrt{5}} \left( + e^{\mathbbm{i} 4 \pi /5} \ket{ \downarrow\uparrow\uparrow\uparrow\uparrow } + e^{\mathbbm{i} 8 \pi /5} \ket{ \uparrow\downarrow\uparrow\uparrow\uparrow } + e^{\mathbbm{i} 2 \pi /5} \ket{ \uparrow\uparrow\downarrow\uparrow\uparrow } + e^{\mathbbm{i} 6 \pi /5} \ket{ \uparrow\uparrow\uparrow\downarrow\uparrow } + \phantom{e^{\mathbbm{i} 0 \pi /5}} \ket{ \uparrow\uparrow\uparrow\uparrow\downarrow } \right)$ 
            \\ 
                $3/2$ & $1/2$  & $4 \pi / 5$                      & $0$                              & --      & --      & --      & $-\frac{\sqrt{5}}{4} \approx -0.5590$      & 
                $\frac{1}{\sqrt{15}} 
                \begin{pmatrix}
                    \phantom{+} \frac{2}{1 + \sqrt{5}} \left( + e^{\mathbbm{i} 6 \pi /5} \ket{ \downarrow\downarrow\uparrow\uparrow\uparrow } + \phantom{e^{\mathbbm{i} 0 \pi /5}} \ket{ \uparrow\downarrow\downarrow\uparrow\uparrow } + e^{\mathbbm{i} 4 \pi /5} \ket{ \uparrow\uparrow\downarrow\downarrow\uparrow } + e^{\mathbbm{i} 8 \pi /5} \ket{ \uparrow\uparrow\uparrow\downarrow\downarrow } + e^{\mathbbm{i} 2 \pi /5} \ket{ \downarrow\uparrow\uparrow\uparrow\downarrow } \right) 
                    \\ 
                    + \frac{1 + \sqrt{5}}{2} \left( + e^{\mathbbm{i} \pi /5 \phantom{1}} \ket{ \uparrow\uparrow\downarrow\uparrow\downarrow } + e^{\mathbbm{i} 5 \pi /5} \ket{ \downarrow\uparrow\uparrow\downarrow\uparrow } + e^{\mathbbm{i} 9 \pi /5} \ket{ \uparrow\downarrow\uparrow\uparrow\downarrow } + e^{\mathbbm{i} 3 \pi /5} \ket{ \downarrow\uparrow\downarrow\uparrow\uparrow } + e^{\mathbbm{i} 7 \pi /5} \ket{ \uparrow\downarrow\uparrow\downarrow\uparrow } \right) 
                    \end{pmatrix}$ 
            \\ 
                $3/2$ & $-1/2$ & $4 \pi / 5$                      & $0$                              & $0$     & --      & --      & $-\frac{\sqrt{5}}{4} \approx -0.5590$      & 
                $\frac{1}{\sqrt{15}} 
                \begin{pmatrix}
                    \phantom{+} \frac{2}{1 + \sqrt{5}} \left( + e^{\mathbbm{i} \pi /5 \phantom{1}} \ket{ \uparrow\uparrow\downarrow\downarrow\downarrow } + e^{\mathbbm{i} 5 \pi /5} \ket{ \downarrow\uparrow\uparrow\downarrow\downarrow } + e^{\mathbbm{i} 9 \pi /5} \ket{ \downarrow\downarrow\uparrow\uparrow\downarrow } + e^{\mathbbm{i} 3 \pi /5} \ket{ \downarrow\downarrow\downarrow\uparrow\uparrow } + e^{\mathbbm{i} 7 \pi /5} \ket{ \uparrow\downarrow\downarrow\downarrow\uparrow } \right) 
                    \\ 
                    + \frac{1 + \sqrt{5}}{2} \left( + e^{\mathbbm{i} 6 \pi /5} \ket{ \downarrow\downarrow\uparrow\downarrow\uparrow } + \phantom{e^{\mathbbm{i} 0 \pi /5}} \ket{ \uparrow\downarrow\downarrow\uparrow\downarrow } + e^{\mathbbm{i} 4 \pi /5} \ket{ \downarrow\uparrow\downarrow\downarrow\uparrow } + e^{\mathbbm{i} 8 \pi /5} \ket{ \uparrow\downarrow\uparrow\downarrow\downarrow } + e^{\mathbbm{i} 2 \pi /5} \ket{ \downarrow\uparrow\downarrow\uparrow\downarrow } \right) 
                    \end{pmatrix}$ 
            \\ 
                $3/2$ & $-3/2$ & $4 \pi / 5$                      & $0$                              & $0$     & $0$     & --      & $-\frac{\sqrt{5}}{4} \approx -0.5590$      & 
                $\frac{1}{\sqrt{5}} \left( + e^{\mathbbm{i} 9 \pi /5} \ket{ \uparrow\downarrow\downarrow\downarrow\downarrow } + e^{\mathbbm{i} 3 \pi /5} \ket{ \downarrow\uparrow\downarrow\downarrow\downarrow } + e^{\mathbbm{i} 7 \pi /5} \ket{ \downarrow\downarrow\uparrow\downarrow\downarrow } + e^{\mathbbm{i} \pi /5 \phantom{1}} \ket{ \downarrow\downarrow\downarrow\uparrow\downarrow } + e^{\mathbbm{i} 5 \pi /5} \ket{ \downarrow\downarrow\downarrow\downarrow\uparrow } \right)$ 
            \\ \hline 
                $3/2$ & $3/2$  & $6 \pi / 5$                      & --                               & --      & --      & --      & $-\frac{\sqrt{5}}{4} \approx -0.5590$      & 
                $\frac{1}{\sqrt{5}} \left( + e^{\mathbbm{i} 6 \pi /5} \ket{ \downarrow\uparrow\uparrow\uparrow\uparrow } + e^{\mathbbm{i} 2 \pi /5} \ket{ \uparrow\downarrow\uparrow\uparrow\uparrow } + e^{\mathbbm{i} 8 \pi /5} \ket{ \uparrow\uparrow\downarrow\uparrow\uparrow } + e^{\mathbbm{i} 4 \pi /5} \ket{ \uparrow\uparrow\uparrow\downarrow\uparrow } + \phantom{e^{\mathbbm{i} 0 \pi /5}} \ket{ \uparrow\uparrow\uparrow\uparrow\downarrow } \right)$ 
            \\ 
                $3/2$ & $1/2$  & $6 \pi / 5$                      & $0$                              & --      & --      & --      & $-\frac{\sqrt{5}}{4} \approx -0.5590$      & 
                $\frac{1}{\sqrt{15}} 
                \begin{pmatrix}
                    \phantom{+} \frac{2}{1 + \sqrt{5}} \left( + e^{\mathbbm{i} 4 \pi /5} \ket{ \downarrow\downarrow\uparrow\uparrow\uparrow } + \phantom{e^{\mathbbm{i} 0 \pi /5}} \ket{ \uparrow\downarrow\downarrow\uparrow\uparrow } + e^{\mathbbm{i} 6 \pi /5} \ket{ \uparrow\uparrow\downarrow\downarrow\uparrow } + e^{\mathbbm{i} 2 \pi /5} \ket{ \uparrow\uparrow\uparrow\downarrow\downarrow } + e^{\mathbbm{i} 8 \pi /5} \ket{ \downarrow\uparrow\uparrow\uparrow\downarrow } \right) 
                    \\ 
                    + \frac{1 + \sqrt{5}}{2} \left( + e^{\mathbbm{i} 9 \pi /5} \ket{ \uparrow\uparrow\downarrow\uparrow\downarrow } + e^{\mathbbm{i} 5 \pi /5} \ket{ \downarrow\uparrow\uparrow\downarrow\uparrow } + e^{\mathbbm{i} \pi /5 \phantom{1}} \ket{ \uparrow\downarrow\uparrow\uparrow\downarrow } + e^{\mathbbm{i} 7 \pi /5} \ket{ \downarrow\uparrow\downarrow\uparrow\uparrow } + e^{\mathbbm{i} 3 \pi /5} \ket{ \uparrow\downarrow\uparrow\downarrow\uparrow } \right) 
                    \end{pmatrix}$ 
            \\ 
                $3/2$ & $-1/2$ & $6 \pi / 5$                      & $0$                              & $0$     & --      & --      & $-\frac{\sqrt{5}}{4} \approx -0.5590$      & 
                $\frac{1}{\sqrt{15}} 
                \begin{pmatrix}
                    \phantom{+} \frac{2}{1 + \sqrt{5}} \left( + e^{\mathbbm{i} 9 \pi /5} \ket{ \uparrow\uparrow\downarrow\downarrow\downarrow } + e^{\mathbbm{i} 5 \pi /5} \ket{ \downarrow\uparrow\uparrow\downarrow\downarrow } + e^{\mathbbm{i} \pi /5 \phantom{1}} \ket{ \downarrow\downarrow\uparrow\uparrow\downarrow } + e^{\mathbbm{i} 7 \pi /5} \ket{ \downarrow\downarrow\downarrow\uparrow\uparrow } + e^{\mathbbm{i} 3 \pi /5} \ket{ \uparrow\downarrow\downarrow\downarrow\uparrow } \right) 
                    \\ 
                    + \frac{1 + \sqrt{5}}{2} \left( + e^{\mathbbm{i} 4 \pi /5} \ket{ \downarrow\downarrow\uparrow\downarrow\uparrow } + \phantom{e^{\mathbbm{i} 0 \pi /5}} \ket{ \uparrow\downarrow\downarrow\uparrow\downarrow } + e^{\mathbbm{i} 6 \pi /5} \ket{ \downarrow\uparrow\downarrow\downarrow\uparrow } + e^{\mathbbm{i} 2 \pi /5} \ket{ \uparrow\downarrow\uparrow\downarrow\downarrow } + e^{\mathbbm{i} 8 \pi /5} \ket{ \downarrow\uparrow\downarrow\uparrow\downarrow } \right) 
                    \end{pmatrix}$ 
            \\ 
                $3/2$ & $-3/2$ & $6 \pi / 5$                      & $0$                              & $0$     & $0$     & --      & $-\frac{\sqrt{5}}{4} \approx -0.5590$      & 
                $\frac{1}{\sqrt{5}} \left( + e^{\mathbbm{i} \pi /5 \phantom{1}} \ket{ \uparrow\downarrow\downarrow\downarrow\downarrow } + e^{\mathbbm{i} 7 \pi /5} \ket{ \downarrow\uparrow\downarrow\downarrow\downarrow } + e^{\mathbbm{i} 3 \pi /5} \ket{ \downarrow\downarrow\uparrow\downarrow\downarrow } + e^{\mathbbm{i} 9 \pi /5} \ket{ \downarrow\downarrow\downarrow\uparrow\downarrow } + e^{\mathbbm{i} 5 \pi /5} \ket{ \downarrow\downarrow\downarrow\downarrow\uparrow } \right)$ 
            \\ \hline 
                $3/2$ & $3/2$  & $8 \pi / 5$                      & --                               & --      & --      & --      & $\frac{\sqrt{5}}{4} \approx 0.5590$        &
                $\frac{1}{\sqrt{5}} \left( + e^{\mathbbm{i} 8 \pi /5} \ket{ \downarrow\uparrow\uparrow\uparrow\uparrow } + e^{\mathbbm{i} 6 \pi /5} \ket{ \uparrow\downarrow\uparrow\uparrow\uparrow } + e^{\mathbbm{i} 4 \pi /5} \ket{ \uparrow\uparrow\downarrow\uparrow\uparrow } + e^{\mathbbm{i} 2 \pi /5} \ket{ \uparrow\uparrow\uparrow\downarrow\uparrow } + \phantom{e^{\mathbbm{i} 0 \pi /5}} \ket{ \uparrow\uparrow\uparrow\uparrow\downarrow } \right)$ 
            \\ 
                $3/2$ & $1/2$  & $8 \pi / 5$                      & $0$                              & --      & --      & --      & $\frac{\sqrt{5}}{4} \approx 0.5590$        & 
                $\frac{1}{\sqrt{15}} 
                \begin{pmatrix}
                    \phantom{+} \frac{1 + \sqrt{5}}{2} \left( + e^{\mathbbm{i} 7 \pi /5} \ket{ \downarrow\downarrow\uparrow\uparrow\uparrow } + e^{\mathbbm{i} 5 \pi /5} \ket{ \uparrow\downarrow\downarrow\uparrow\uparrow } + e^{\mathbbm{i} 3 \pi /5} \ket{ \uparrow\uparrow\downarrow\downarrow\uparrow } + e^{\mathbbm{i} \pi /5 \phantom{1}} \ket{ \uparrow\uparrow\uparrow\downarrow\downarrow } + e^{\mathbbm{i} 9 \pi /5} \ket{ \downarrow\uparrow\uparrow\uparrow\downarrow } \right) 
                    \\ 
                    + \frac{2}{1 + \sqrt{5}} \left( + e^{\mathbbm{i} 2 \pi /5} \ket{ \uparrow\uparrow\downarrow\uparrow\downarrow } + \phantom{e^{\mathbbm{i} 0 \pi /5}} \ket{ \downarrow\uparrow\uparrow\downarrow\uparrow } + e^{\mathbbm{i} 8 \pi /5} \ket{ \uparrow\downarrow\uparrow\uparrow\downarrow } + e^{\mathbbm{i} 6 \pi /5} \ket{ \downarrow\uparrow\downarrow\uparrow\uparrow } + e^{\mathbbm{i} 4 \pi /5} \ket{ \uparrow\downarrow\uparrow\downarrow\uparrow } \right) 
                    \end{pmatrix}$ 
            \\ 
                $3/2$ & $-1/2$ & $8 \pi / 5$                      & $0$                              & $0$     & --      & --      & $\frac{\sqrt{5}}{4} \approx 0.5590$        & 
                $\frac{1}{\sqrt{15}} 
                \begin{pmatrix}
                    \phantom{+} \frac{1 + \sqrt{5}}{2} \left( + e^{\mathbbm{i} 2 \pi /5} \ket{ \uparrow\uparrow\downarrow\downarrow\downarrow } + \phantom{e^{\mathbbm{i} 0 \pi /5}} \ket{ \downarrow\uparrow\uparrow\downarrow\downarrow } + e^{\mathbbm{i} 8 \pi /5} \ket{ \downarrow\downarrow\uparrow\uparrow\downarrow } + e^{\mathbbm{i} 6 \pi /5} \ket{ \downarrow\downarrow\downarrow\uparrow\uparrow } + e^{\mathbbm{i} 4 \pi /5} \ket{ \uparrow\downarrow\downarrow\downarrow\uparrow } \right) 
                    \\ 
                    + \frac{2}{1 + \sqrt{5}} \left( + e^{\mathbbm{i} 7 \pi /5} \ket{ \downarrow\downarrow\uparrow\downarrow\uparrow } + e^{\mathbbm{i} 5 \pi /5} \ket{ \uparrow\downarrow\downarrow\uparrow\downarrow } + e^{\mathbbm{i} 3 \pi /5} \ket{ \downarrow\uparrow\downarrow\downarrow\uparrow } + e^{\mathbbm{i} \pi /5 \phantom{1}} \ket{ \uparrow\downarrow\uparrow\downarrow\downarrow } + e^{\mathbbm{i} 9 \pi /5} \ket{ \downarrow\uparrow\downarrow\uparrow\downarrow } \right) 
                    \end{pmatrix}$ 
            \\ 
                $3/2$ & $-3/2$ & $8 \pi / 5$                      & $0$                              & $0$     & $0$     & --      & $\frac{\sqrt{5}}{4} \approx 0.5590$        & 
                $\frac{1}{\sqrt{5}} \left( + e^{\mathbbm{i} 3 \pi /5} \ket{ \uparrow\downarrow\downarrow\downarrow\downarrow } + e^{\mathbbm{i} \pi /5 \phantom{1}} \ket{ \downarrow\uparrow\downarrow\downarrow\downarrow } + e^{\mathbbm{i} 9 \pi /5} \ket{ \downarrow\downarrow\uparrow\downarrow\downarrow } + e^{\mathbbm{i} 7 \pi /5} \ket{ \downarrow\downarrow\downarrow\uparrow\downarrow } + e^{\mathbbm{i} 5 \pi /5} \ket{ \downarrow\downarrow\downarrow\downarrow\uparrow } \right)$ 
            \\ \hline 
                $5/2$ & $5/2$  & --                               & --                               & --      & --      & --      & $\frac{5}{4} = 1.25$                       & 
                $\phantom{+\frac{1}{\sqrt{1}}}$ $\ket{ \uparrow\uparrow\uparrow\uparrow\uparrow }$ 
            \\ 
                $5/2$ & $3/2$  & $0$                              & --                               & --      & --      & --      & $\frac{5}{4} = 1.25$                       & 
                $\frac{1}{\sqrt{5}} \left( + \ket{ \downarrow\uparrow\uparrow\uparrow\uparrow } + \ket{ \uparrow\downarrow\uparrow\uparrow\uparrow } + \ket{ \uparrow\uparrow\downarrow\uparrow\uparrow } + \ket{ \uparrow\uparrow\uparrow\downarrow\uparrow } + \ket{ \uparrow\uparrow\uparrow\uparrow\downarrow } \right)$ 
            \\ 
                $5/2$ & $1/2$  & $0$                             & $0$                               & --      & --      & --      & $\frac{5}{4} = 1.25$                       & 
                $\frac{1}{\sqrt{10}} 
                \begin{pmatrix} 
                    + \ket{ \downarrow\downarrow\uparrow\uparrow\uparrow } + \ket{ \uparrow\downarrow\downarrow\uparrow\uparrow } + \ket{ \uparrow\uparrow\downarrow\downarrow\uparrow } + \ket{ \uparrow\uparrow\uparrow\downarrow\downarrow } + \ket{ \downarrow\uparrow\uparrow\uparrow\downarrow } 
                    \\
                    + \ket{ \uparrow\uparrow\downarrow\uparrow\downarrow } + \ket{ \downarrow\uparrow\uparrow\downarrow\uparrow } + \ket{ \uparrow\downarrow\uparrow\uparrow\downarrow } + \ket{ \downarrow\uparrow\downarrow\uparrow\uparrow } + \ket{ \uparrow\downarrow\uparrow\downarrow\uparrow } 
                \end{pmatrix}$ 
            \\ 
                $5/2$ & $-1/2$ & $0$                             & $0$                               & $0$     & --      & --      & $\frac{5}{4} = 1.25$                       & 
                $\frac{1}{\sqrt{10}} 
                \begin{pmatrix}
                    + \ket{ \uparrow\uparrow\downarrow\downarrow\downarrow } + \ket{ \downarrow\uparrow\uparrow\downarrow\downarrow } + \ket{ \downarrow\downarrow\uparrow\uparrow\downarrow } + \ket{ \downarrow\downarrow\downarrow\uparrow\uparrow } + \ket{ \uparrow\downarrow\downarrow\downarrow\uparrow } 
                    \\
                    + \ket{ \downarrow\downarrow\uparrow\downarrow\uparrow } + \ket{ \uparrow\downarrow\downarrow\uparrow\downarrow } + \ket{ \downarrow\uparrow\downarrow\downarrow\uparrow } + \ket{ \uparrow\downarrow\uparrow\downarrow\downarrow } + \ket{ \downarrow\uparrow\downarrow\uparrow\downarrow } 
                \end{pmatrix}$ 
            \\ 
                $5/2$ & $-3/2$ & $0$                             & $0$                               & $0$     & $0$     & --      & $\frac{5}{4} = 1.25$                       & 
                $\frac{1}{\sqrt{5}} \left( + \ket{ \uparrow\downarrow\downarrow\downarrow\downarrow } + \ket{ \downarrow\uparrow\downarrow\downarrow\downarrow } + \ket{ \downarrow\downarrow\uparrow\downarrow\downarrow } + \ket{ \downarrow\downarrow\downarrow\uparrow\downarrow } + \ket{ \downarrow\downarrow\downarrow\downarrow\uparrow } \right)$ 
            \\ 
                $5/2$ & $-5/2$ & $0$                             & $0$                               & $0$     & $0$     & $0$     & $\frac{5}{4} = 1.25$                       & 
                $\phantom{+\frac{1}{\sqrt{1}}}$ $\ket{ \downarrow\downarrow\downarrow\downarrow\downarrow }$ 
            \\ \hline 
        \end{tabular}
        \\
        \vspace{1em}
        \raggedright
        The numerical values in the table admit exact expressions: 
        \qquad \quad
        $1.1989 \approx \operatorname{ArcCoth}[\sqrt{\frac{5}{19}\left(1 + 2\sqrt{5}\right)}]$, 
        \\
        $1.7053 \approx \pi - \operatorname{ArcTan}[\sqrt{13 + \frac{32}{\sqrt{5}} + 2 \sqrt{93 + \frac{208}{\sqrt{5}}}}]$, 
        \quad
        $3.3212 \approx \pi + \operatorname{ArcTan}[\sqrt{13 + \frac{32}{\sqrt{5}} - 2 \sqrt{93 + \frac{208}{\sqrt{5}}}}]$, 
        \\
        $2.9620 \approx \pi - \operatorname{ArcTan}[\sqrt{13 + \frac{32}{\sqrt{5}} - 2 \sqrt{93 + \frac{208}{\sqrt{5}}}}]$, 
        \quad
        $4.5779 \approx \pi + \operatorname{ArcTan}[\sqrt{13 + \frac{32}{\sqrt{5}} + 2 \sqrt{93 + \frac{208}{\sqrt{5}}}}]$. 

        \end{table}
    }
    
    \clearpage

%=============================================================================
\end{document}